\documentclass[]{ar2e-changed}
\usepackage{graphicx}        
\usepackage{bm}

\begin{document}



\newcommand{\beq}{\begin{equation}}
\newcommand{\eeq}{\end{equation}}
\newcommand{\beqa}[0]{\begin{eqnarray}}
\newcommand{\eeqa}[0]{\end{eqnarray}}
\newcommand{\kf}{k_{\rm F}}
\newcommand{\vf}{v_{\rm F}}
\newcommand{\Ef}{E_{\rm F}}
\newcommand{\ef}{\varepsilon_{\rm F}}
\newcommand{\nab}{\overrightarrow{\nabla}}
\newcommand{\galnab}{\stackrel{\leftrightarrow}{\nabla}}
\newcommand{\vlowk}{V_{{\rm low}\,k}}
\newcommand{\bfx}[0]{{\bf x}}
\newcommand{\Vext}{v_{\rm ext}}
\renewcommand{\vec}[1]{{\bf #1}}
\newcommand{\cdotb}{\bm{\cdot}}


\title{Effective Field Theory and Finite Density Systems}

\markboth{Furnstahl, Rupak \& Sch\"afer}{EFT and Finite Density Systems}

\author{Richard J. Furnstahl
\affiliation{Department of Physics, Ohio State University;
  email: furnstahl.1@osu.edu}
Gautam Rupak and Thomas Sch\"afer
\affiliation{Department of Physics, North Carolina State University;
 email: grupak@gmail.com, tmschaef@ncsu.edu}}


\begin{abstract}
This review gives an overview of effective field theory (EFT) as 
applied at finite density, with a focus on nuclear many-body systems.
Uniform systems with short-range interactions illustrate the 
ingredients and virtues of many-body EFT and then the varied frontiers 
of EFT for finite nuclei and nuclear matter are surveyed.   
\end{abstract}

\maketitle


\section{INTRODUCTION}
  \label{sec:intro}

Calculating the properties of atomic nuclei and nuclear matter starting from
microscopic internucleon forces
is one of the oldest unsolved challenges of nuclear physics.
Renewed interest in this problem is fueled by experiments
at rare isotope facilities, which open the door to new domains of 
unstable nuclides that are not all accessible in the lab, 
and by descriptions of astrophysical phenomena such as supernovae and neutron
stars, which require controlled extrapolations of the equation of state 
of nuclear matter in density, temperature, and proton
fraction~\cite{NAP:1999}.
But despite decades of work and technological advances,
there remain severe computational barriers and 
only limited control of uncertainties in conventional nuclear many-body
calculations of all but the lightest nuclei.
The difficulties are
exacerbated by the need to supplement accurate phenomenological
two-nucleon potentials with
poorly understood many-body forces to achieve a quantitative 
(and in many cases qualitative) description of nuclei.
Finally, conventional approaches are at best loosely connected to
quantum chromodynamics (QCD), the underlying theory of the strong interaction.

Effective field theory (EFT) provides new tools
to address these challenges.
The goals of EFT applied to finite density nuclear systems are to put
nuclear many-body physics on a firm foundation so that
it can be \textit{i)} systematically improved with associated theoretical 
error bars,
\textit{ii)} extended reliably to regimes where there is limited 
or no data,
and \textit{iii)} connected to QCD as well as to few-body experiments.
Our modest aim for this review is to give a flavor for how EFT 
can accomplish these goals in many-body systems 
and to survey the frontiers of EFT-based calculations of 
many-body nuclei and nuclear matter.

Any EFT builds on
a basic physics principle that underlies \emph{every} low-energy
effective model or theory.
A high-energy, short-wavelength probe sees details down to scales comparable
to the wavelength.  Thus, 
electron scattering at sufficiently high energy reveals the quark substructure
of protons and neutrons in a nucleus. 
But at lower energies, details are not resolved,
and one can replace short-distance structure with something simpler, as in
a multipole expansion of a complicated charge or current distribution.
This means it is not necessary to calculate with 
full QCD to do accurate strong interaction physics
at low energies; we can replace quarks and gluons by neutrons
and protons (and maybe pions and \ldots).
Effective field theory provides a systematic, model-independent way 
to carry out this program starting with a local Lagrangian framework.

An EFT is formulated by specifying appropriate low-energy degrees of freedom
and then constructing the Lagrangian as a complete set of terms
that embody the symmetries of the underlying theory.
(Note: the general Lagrangian will typically be \emph{overcomplete},
but redundant terms can be removed by redefining the fields appropriately.)
There is not a unique EFT for nuclear physics.
In different applications the relevant degrees of freedom might be
neutrons and protons only, or neutrons, protons, and pions, or 
neutrons, protons, pions, and $\Delta$'s or quasi-nucleons.
The form of the EFT can be chosen to 
readily expose universal behavior, such as features
dilute neutron matter has in
common with phenomena seen in cold atom experiments.

In applying an EFT Lagrangian, one must 
confront in a controlled way
the impact of excluded short-distance physics.
Quantum mechanics implies that
sensitivity to short-distance physics is \emph{always} present in a
low-energy theory, but it
is made manifest in an EFT through dependence on a cutoff or other
regulator instead of being hidden in phenomenological form factors.
Removing this dependence necessitates 
a well-defined regularization and renormalization 
scheme as part of the EFT specification.
This necessity becomes a virtue as residual regulator dependence
can be used to assess truncation errors and many-body approximations.
Furthermore, the freedom in how to regulate coupled with the freedom to make 
unitary transformations
can be exploited by renormalization group methods to greatly simplify
few- and many-body nuclear calculations.

For an EFT calculation to be improvable order-by-order,
one needs a scheme to organize
the infinity of possible terms in the
Lagrangian based on an expansion
parameter (or parameters).
Such a scheme  is called a power counting.
Power counting tells us what terms (or Feynman diagrams)
to include at each order and lets us estimate
the theoretical truncation error.  
The radius of convergence associated with the expansion means that the EFT
predicts it own downfall, in contrast to phenomenological models.
EFT expansion parameters most commonly arise as a ratio of disparate
physical scales rather than as a small coupling constant (\textit{e.g.}, as in
Coulomb systems); a many-body example is the ratio of 
the range of the interaction to the interparticle
spacing in a dilute system. 
The power counting for this example is particularly simple
when the scattering length is roughly the same size as the interaction range 
(called ``natural'') but changes
dramatically if the scattering length is much larger (called ``unnatural'').
We will explore both situations below.

Chiral effective field theory is a faithful low-energy realization
of QCD whose power counting takes advantage of the spontaneously
broken chiral symmetry that gives rise to the almost massless (on
hadronic scales) pion.
It has the potential to bridge the gap between QCD and nuclei,
letting us explore how nuclear properties depend
on QCD parameters 
(\textit{e.g.}, how would
the binding energies of nuclei change if the light quark masses
were different or if the QCD scale parameter were time dependent?)
and opening a connection to \textit{ab initio} QCD lattice
calculations.
Chiral EFT power counting explains the empirical hierarchy of many-body forces
in nuclear physics, fixes their natural sizes,
and gives
an organizing principle for their construction.
Other compelling features are the systematic inclusion
of relativistic corrections and prescriptions for consistent
currents needed to predict experimental observables.

It is probably evident that
a comprehensive treatment of  EFT and finite density nuclear systems
would require several extended reviews covering EFT in general,
EFT applied to internucleon interactions, 
and field theory at finite density.  
Fortunately there are recent articles in this journal
to provide much of the background for the interested reader;
these include an introduction to 
effective field theory by Burgess~\cite{Burgess:2007pt}, 
an overview of chiral perturbation theory by 
Bernard and Mei{\ss}ner~\cite{Bernard:2006gx},
and a review of EFT for few-nucleon systems by Bedaque and van
Kolck~\cite{Bedaque:2002mn} (see also refs.~\cite{Beane:2000fx,Epelbaum:2005pn}).
We will focus here on illustrating how the basic principles
of EFT can be realized at finite density
and on surveying various applications to nuclear matter and finite nuclei. 
Our treatment will be schematic in most cases
and we will refer the reader to the literature for details.

In Section~\ref{sec:uniform},  we consider uniform systems with short-range
interactions. The dilute Fermi gas with repulsive interactions serves as a
prototype for EFT at finite density
while new features and techniques arise when we
study physics near the Fermi surface.  Many-body systems with unnatural
scattering lengths, which manifest various forms of
universal physics, are attacked by a
variety of nonperturbative EFT techniques. 
Actual applications of EFT to nuclear
many-body systems are in their infancy and there are multiple frontiers; 
a range of examples are described in Section~\ref{sec:nuclei}. These start with
the use of chiral EFT interactions as input to conventional many-body wave
function methods applicable to light nuclei and a pioneering attempt to apply EFT
to the methods themselves. Lattice calculations provide a
complementary nonperturbative approach. Perturbative chiral EFT calculations
for nuclei may be possible, however, if the power counting differs at
nuclear densities. This may be justified by renormalization group
transformations that soften the chiral interactions.  Finally, density
functional theory (DFT), 
which is computationally tractable for all nuclides, is
naturally cast in EFT form using effective actions. We conclude in
Section~\ref{sec:summary} with a summary of the current status of EFT for nuclear
systems, on-going developments, important open questions, and
pointers to omitted topics.


\section{EFT FOR UNIFORM SYSTEMS} 
 \label{sec:uniform}

In this section, we illustrate the ideas of EFT at finite density
for uniform systems with short-range interactions.

\subsection{Prototype Many-Body EFT}
 \label{subsec:prototype}

We start with perhaps the simplest possible application, a dilute
Fermi system with repulsive, spin-independent interactions of range $R$.
A concrete example would be ``hard sphere'' repulsion at radius $R$,
which can be viewed as a caricature of the short-range part of
the nuclear force.
In perturbation theory all matrix elements of this potential are
infinite; while a more realistic potential would not be so extreme,
textbook treatments of this many-body problem all start with nonperturbative
summations and then expansions at low-density~\cite{FETTER71}.
In contrast, the EFT approach directly exploits the essential physics 
that the ``hard core'' is not resolved at low momentum. 

With either approach, the end result for
free space, two-particle scattering at low energies 
($\lambda = 2\pi/k \gg 1/R$) 
is the effective range expansion;
\textit{e.g.}, the $s$-wave phase
shift $\delta_0(k)$ satisfies:
\beq k \cot \delta_0(k) \stackrel{k\rightarrow 0}{\longrightarrow}
   -\frac{1}{a_0} + \frac12 r_0 k^2 + \ldots
   \label{eq:ERE}
\eeq 
where
${a_0}$ is the scattering length and $r_0$ is the effective range.
The system is said to have a ``natural'' scattering length if
it is the same order as the range of the interaction (\textit{e.g.}, $a_0 = R$ and
$r_0 = 2R/3$ for hard spheres).
In a later section we consider the case of unnatural scattering length,
with $a_0 \gg R$, which is
relevant for dilute neutron matter and cold atom systems.
For a natural system, the dilute expansion of the energy density
for a uniform system starts as ($s$-wave only here)
\beq
 {\cal E} =
    { \rho} \frac{{ \kf^2}}{2M}
     \biggl[\frac{3}{5} + 
     (\nu-1)\biggl\{{\frac{2}{3\pi}}{ (\kf a_0)}
     +
       { \frac{4}{35\pi^2}(11-2\ln 2)}{ (\kf a_0)^2}
    + \frac{1}{10\pi}{ (\kf r_0)(\kf a_0)^2}\biggr\}
       + \cdots
     \biggr] \;,
     \label{eq:Edensity}  
\eeq
where $\kf$ is the Fermi momentum, $\nu$ is the spin degeneracy,
and $\rho = \nu\kf^3/6\pi^2$.  
This result arises very cleanly from an EFT
treatment~\cite{Hammer:2000xg}.  

Consider
the ingredients for any effective field theory along with 
the specifics for this example:
  \begin{enumerate}
    \item \emph{Use the most general ${\cal L}$ with low-energy 
     degrees-of-freedom consistent
      with  global and local symmetries of the underlying theory.} 
    Here we have nucleons only with Galilean invariance and
    discrete symmetries.  A general interaction is then
     a sum of delta functions and derivatives of delta functions
     with two-body (four fields), three-body (six-fields), and so on,
     so ${\cal L}_{\rm eft}$ is
   \beq
      {\cal L}_{\rm eft} = 
         \psi^\dagger \bigl[i\frac{\partial}{\partial t} 
         + \frac{\nabla^{\,2}}{2M}\bigr]
           \psi - \frac{{ C_0}}{2}(\psi^\dagger \psi)^2
            + \frac{{ C_2}}{16}\bigl[ (\psi\psi)^\dagger 
                                  (\psi\!\galnab\!{}^2\psi)+\mbox{ h.c.} 
                             \bigr]
              - \frac{{D_0}}{6}(\psi^\dagger \psi)^3 +  \ldots \;,
	   \label{eq:Lagrangian}   
   \eeq
   where $\ldots$ indicates terms with more derivatives and more fields.
   (We have eliminated higher-order time derivatives using the
   equations of motion~\cite{Hammer:2000xg}.)
   The $\psi$'s have $\nu$ components and spin-indices are implicit
   (and contracted between $\psi^\dagger$ and $\psi$).

   \item  \emph{Declare a regularization and renormalization scheme.}
     One choice is to smear out the delta functions (\textit{e.g.}, 
     as gaussians in
     momentum space) to introduce a cutoff; renormalization
     would remove cutoff dependence.
     However, for a natural $a_0$, using
     dimensional regularization and minimal subtraction (rather
     than a cutoff)
     is particularly convenient and efficient.	   

   \item \emph{Establish a well-defined power counting}, which means
     identifying small expansion parameters, typically
     using a ratio of scales.
     In free space $k/\Lambda$ 
     with $\Lambda \sim 1/R$ is the clear choice, and then
     $\kf/\Lambda$ is the corresponding parameter in the medium.
     Dimensional analysis, with some additional insight to
     give us the $4\pi$'s, implies ($2i$ denotes the number of 
     gradients)
    \beq
      C_{2i} \sim \frac{\textstyle 4\pi}{\textstyle M } R^{2i+1}\; ,
      \quad
       D_{2i} \sim 
      \frac{\textstyle 4\pi}{\textstyle M} R^{2i+4}
      \; , 
      \label{eq:scaling}
    \eeq   
    which will enable us to make quantitative power-counting estimates.      
  \end{enumerate} 
Feynman diagrams and rules for the EFT follow from conventional formalism
for free-space and many-body perturbation theory (\textit{e.g.}, 
see \cite{FETTER71,Negele:1988vy}).

The constants $C_{2i}$ are determined by matching
to the free-space scattering amplitude $f_0(k)$ in perturbation theory,
 \beq 
    {f_0(k)} =
   \frac{4\pi}{M}\left(
      a_0  -i a_0^2 k - a_0^3 k^2 + a_0^2 r_0 k^2 + \cdots
     \right) \; .
     \label{eq:scatt}
 \eeq
The leading potential $V^{(0)}_{\rm EFT}(\bfx) = C_0 \delta(\bfx)$ or
$\langle {\bf k} | V^{(0)}_{\rm eft} | {\bf k'}\rangle = C_0$,
where ${\bf k}, {\bf k'}$ are relative momenta.
Matching to $f_0(k)$ fixes $C_0 = 4\pi a_0/M$ at leading order, 
which then determines
the leading finite density contribution (Hartree-Fock) 
in Eq.~(\ref{eq:Edensity}) after sums over the Fermi sea:
 \beq
   \raisebox{-.19in}{\includegraphics*[width=.4in,angle=0]{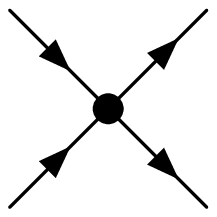}}
   \longrightarrow \  C_0
    {\quad \Longrightarrow \quad}
   \raisebox{-.08in}{%
  \includegraphics*[width=0.7in,angle=0]{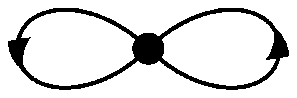}}
  \ \longrightarrow \ {\cal E}_{\rm LO} 
     = \frac{C_0}{2}\nu(\nu-1)\left(\sum_{\bf k}^{\kf} 1\right)^2
  \ { \propto a_0 \kf^6} \; .
 \eeq
Similar matching yields $C_2$ in
terms of $a_0$ and $r_0$ and the corresponding
Hartree-Fock contribution for the effective range.

At the next order is 
$\langle {\bf k} | V^{(0)}_{\rm eft} G_0 V^{(0)}_{\rm eft} | {\bf k'}\rangle$,
which includes a linearly divergent loop integral:
 \beq
  \raisebox{-.2in}{\includegraphics*[width=0.7in,angle=0]{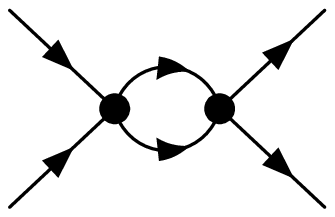}}
   \ \longrightarrow\ 
  C_0 M \int^{\Lambda_c}\! \frac{d^3q}{(2\pi)^3} \frac{1}{k^2-q^2 + i\epsilon}\, C_0
  =
  C_0^2 M \left(
  \frac{\Lambda_c}{2\pi^2}
       - \frac{ik}{4\pi} + {\cal O}(\frac{k^2}{\Lambda_c})
    \right)   \; . \eeq
We can redefine (``renormalize'') $C_0$ to absorb the linear
dependence on the cutoff $\Lambda_c$, but we'll have higher powers of $k$
from every diagram.
A more efficient scheme is 
dimensional regularization with minimal subtraction (DR/MS), which implies 
only one power of $k$ survives:
 \beq
   \int\! \frac{d^Dq}{(2\pi)^3} \frac{1}{k^2-q^2 + i\epsilon}
     \stackrel{D\rightarrow 3}{\longrightarrow} 
       - \frac{ik}{4\pi} \; . 
 \eeq
Then we get the second term in Eq.~(\ref{eq:scatt}) automatically with no
change in $C_0$.  At higher orders there is 
exactly one power of $k$ per diagram and \emph{natural} coefficients
[\textit{i.e.}, consistent with Eq.~(\ref{eq:scaling})],
so we can
estimate truncation errors from simple dimensional analysis.

The contribution to the energy density has two terms, one of which
vanishes identically.  In the other,
we get a linear divergence again,
 \beq
  \raisebox{-.18in}{\includegraphics*[width=0.7in,angle=0]{fig_C0sq}}
 {\quad \Longrightarrow \quad}
  \raisebox{-.2in}{%
  \includegraphics*[width=.4in,angle=0]{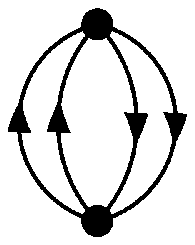}}
  \ \longrightarrow \ {\cal E}_{\rm NLO} \propto
         \int_{\kf}^\infty\! \frac{d^3q}{(2\pi)^3} \frac{C_0^2}{k^2-q^2}
	 \; ,
 \eeq
but the \emph{same} renormalization fixes it,  
 \beq
         {\int_{\kf}^\infty\! \frac{1}{k^2-q^2}}
     =     
       { \int_{0}^\infty\!  \frac{1}{k^2-q^2}}
      { - \int_{0}^{\kf}\!  \frac{1}{k^2-q^2}}
    \stackrel{ D\rightarrow 3}{\longrightarrow}
     { - \int_{0}^{\kf}\!  \frac{1}{k^2-q^2}}
   \;,
 \eeq
and particles become
holes through the renormalization.
Pauli blocking doesn't change the free-space ultraviolet (short-distance)
renormalization, since the density is a long-distance effect;
after fixing free space, the in-medium renormalization is determined.
We find ${\cal E}_{\rm NLO} \propto a_0^2 \kf^7$.

The diagrammatic power counting with DR/MS is very simple, with each
loop adding a power of $k$ in free-space.
At finite density, a diagram with $V^n_{2i}$ $n$--body vertices 
and $2i$ gradients scales as ${ (\kf)^\beta}$ with
 \beq
  \beta=5+\sum_{n=2}^\infty \sum_{i=0}^\infty (3n+2i-5) V_{2i}^n \;.
  \label{eq:naturalpc}
 \eeq
This reproduces, for example, the leading order 
$[\beta = 5 + (3\cdot 2 + 2\cdot 0 - 5)\cdot 1  = 6]$ and
next-to-leading order 
$[\beta = 5 + (3\cdot 2 + 2\cdot 0 - 5)\cdot 2  = 7]$ dependencies.
The power counting is exceptionally clean, with a
separation of vertex factors $\propto a_0,r_0,\ldots$ and
a dimensionless geometric integral times $\kf^\beta$,
with each diagram contributing to exactly one order in the
expansion.
There is a systematic hierarchy, since 
adding derivatives or higher-body interactions increases the power of $\kf$. 	 
The ratio of successive terms is $\sim \kf R$ [\textit{e.g.}, in
Eq.~(\ref{eq:Edensity})], so we can 
estimate excluded contributions. 

The energy density (\ref{eq:Edensity})
\emph{looks} like a power series in $\kf$, but at higher order
there are \emph{logarithmic} divergences from 3--3 scattering,
which indicate new sensitivity to short-distance behavior.
A cutoff $\Lambda_c$ serves as a resolution scale;  as we
increase $\Lambda_c$, we see more of the short-distance details.
Observables (such as scattering amplitudes) must not vary
with $\Lambda_c$, so changes must be absorbed in a coupling.
But it can't be a coupling from 2--2 scattering, because
we already took care of all the divergences there.
We instead must use the point-like three-body force,
whose coupling $D_0(\Lambda_c)$ can absorb the dependence
on $\Lambda_c$~\cite{BRAATEN97}.
The diagrams are $\propto { (C_0)^4 \ln(k/\Lambda_c)}$, which means
 \beq
   \frac{d}{d\Lambda_c}\biggl[
   \raisebox{-.2in}{%
      \includegraphics*[width=2.4in,angle=0.0]{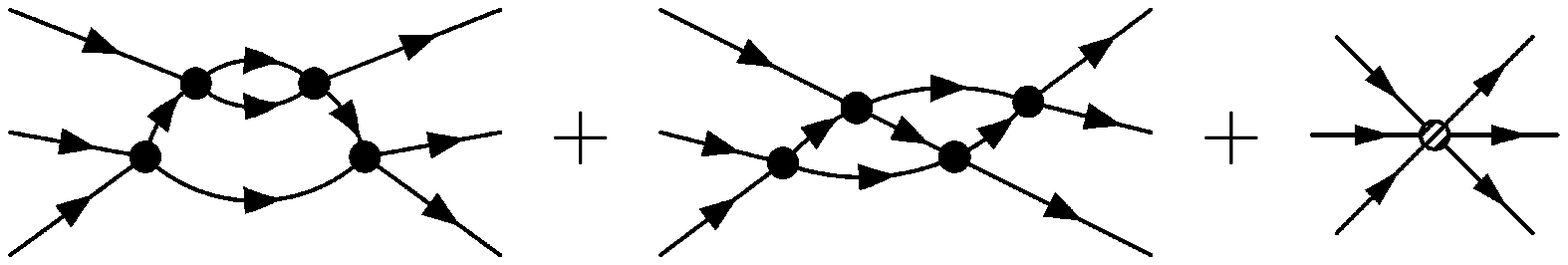}}
   \biggr] = 0
 \ \Longrightarrow\
    D_0(\Lambda_c) \propto { (C_0)^4 \ln (a_0\Lambda_c)}
 \eeq
fixes the coefficient $D_0(\Lambda_c)$.
Dimensional regularization is similar~\cite{Hammer:2000xg}.
In turn this implies for the energy density,
  \beq 
    \raisebox{-.25in}{\includegraphics*[width=2.8in]{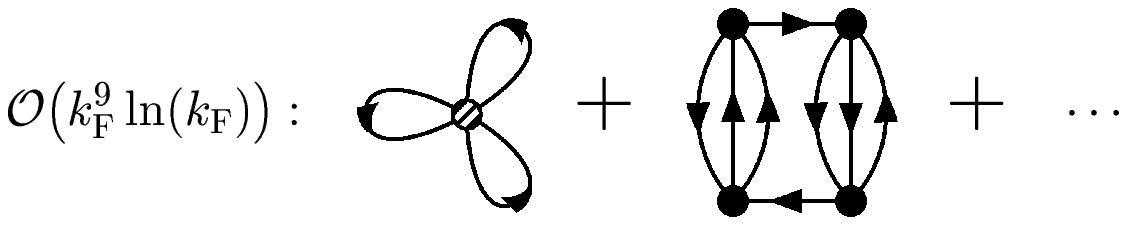}}       
    \propto (\nu-2)(\nu-1)
    \kf^5 (\kf a_0)^{4} \, {\ln(\kf a_0)}
  \eeq
without actually carrying out the calculation!
Similar analyses can identify the higher logarithmic terms in the
expansion of the energy density~\cite{BRAATEN97,Hammer:2000xg}. 
This is an example of the inevitability of many-body forces in
low-energy theories: when the resolution or degrees of freedom are
changed, we will have many-body forces.
Thus the question is not whether such forces are 
present, but how large they are.  For nuclear physics, their natural
size implies they cannot be neglected.
 
This brief tour
of the EFT for a natural dilute Fermi gas included features common
to many other applications.
Even if we knew that the underlying physics  was a hard-sphere potential,
the EFT was easier to calculate 
than conventional approaches~\cite{FETTER71}.
Further, the EFT directly reveals the universal
nature of the many-body counterpart to the effective range expansion,
which applies to any short-range repulsive potential. 
Of course, this example is very  simple; there are many ways
to generalize.
Some are immediate: \textit{e.g.},
we can account for short-range spin-dependent interactions by
adding terms such as 
$C^\sigma_0(\psi^\dagger\bm{\sigma}\psi)\bm{\cdot}
      (\psi^\dagger\bm{\sigma}\psi)$.
If we consider unnatural scattering, however, then we must revisit
the power counting and consider alternative expansion parameters
since $\kf a_0$ is no longer small.
But first we turn from EFT for  bulk properties to EFT near the Fermi surface.

\subsection{EFT Near the Fermi Surface}
\label{sec_fl}

 The theory described in the last section is completely perturbative. 
At any order in the $\kf R$ expansion only a finite number of diagrams 
has to be computed. There are two ways in which this expansion can fail. 
One possibility is that one of the effective range parameters (typically,
the scattering length) is anomalously large, so that a certain class of 
diagrams has to be summed to all orders. We will study this problem 
in Section~\ref{sec:unitary}. A second possibility is that the density 
(and the Fermi momentum) is too large and $\kf R$ ceases to be a useful 
expansion parameter. In this case it is possible to construct a different 
kind of effective field theory by focusing on quasi-particles in the 
vicinity of the Fermi surface, and using $|k-\kf |/\Lambda$ as an 
expansion parameter. This effective theory is known as Landau Fermi 
liquid theory \cite{Pines:1966,Baym:1991}. The Landau theory does
not account for all properties of the many-body system, but it does
describe phenomena that are sensitive to physics near the Fermi 
surface such as collective modes, pairing, transport properties, 
\textit{etc}.

 Fermi liquid theory was originally developed by Landau using intuitive 
arguments. These arguments were later confirmed by Abrikosov and others 
using diagrammatic many-body perturbation theory \cite{Abrikosov:1963}.
The modern view of Fermi liquid theory as an effective field theory
was advocated by Shankar, Polchinski, and others 
\cite{Shankar:1993pf,Polchinski:1992ed}. Consider the effective action 
of non-interacting, nonrelativistic Fermions near a Fermi surface
\beq 
\label{seff_fs}
  S = \int\! dt\, \int\! \frac{d^3p}{(2\pi)^3}\,
 \psi^\dagger(p)\left(i\frac{\partial}{\partial t} -\vf l_p\right)\psi(p)
 \;. 
\eeq
Here we have decomposed the momenta as $\vec{p}=\vec{k}+\vec{l}_p$,
where $\vec{k}$ is on the Fermi surface, $|\vec{k}|=\kf $, and 
$\vec{l}_p$ is orthogonal to the Fermi surface. The Fermi velocity 
is defined as $\vf =\partial E_p/\partial p$, where $E_p$ is the 
quasiparticle energy. The power counting can be established by studying 
the behavior of operators under transformations $l_p\to sl_p$ that 
scale the momenta towards the Fermi surface. Writing $E_p=\Ef +\vf l_p
+{\cal O}(l_p^2)$ we see that as $s\to 0$ only the Fermi velocity survives, 
so the detailed form of the dispersion relation is irrelevant. Using 
$d^3p=\kf ^2(dl_p)(d\Omega)$ we observe that $d^3p\sim s$, $dt\sim 
s^{-1}$, and $\psi\sim s^{-1/2}$ and $S$ in (\ref{seff_fs}) is 
${\cal O}(s^0)$. 

We can now study the importance 
of interactions between fermions near the Fermi surface. The most 
general four-fermion interaction is of the form
\beq
\label{sfl_4f}
S_{4f} = \frac{1}{4}\int\! dt \left[ \prod_{i=1}^{4}\int\!
     \frac{d^3p_i}{(2\pi)^3}\right]
     \psi^\dagger(\vec{p}_4)\psi^\dagger(\vec{p}_3)
     \psi(\vec{p}_2)\psi(\vec{p}_1) \delta^{3}\left(\vec{p}_{\rm tot}\right) 
     U(\vec{p}_4,\vec{p}_3,\vec{p}_2,\vec{p}_1) \;,
\eeq
where $\vec{p}_{\rm tot}$ is the sum of the four momenta $\vec{p}_i$, and
we have suppressed the spin labels on $U$. For a generic set of momenta 
$\vec{p}_i$ the delta function constrains the large components of the 
momenta and scales as $\delta^3(\vec{p}_{\rm tot})\sim s^0$. In this case 
the four-fermion interaction scales as $s^1$ and becomes irrelevant 
near the Fermi surface. Interactions involving more fermions are 
even more strongly suppressed. 

 An exception occurs if the large components of the momenta cancel. 
This happens for back-to-back momenta, $\vec{k}_1=-\vec{k}_2$, and
for generalized forward scattering, $\vec{k_1}\cdotb\vec{k}_2= 
\vec{k}_3\cdotb\vec{k}_4$. In these cases
one component of the delta functions constrains $\vec{l}$, the 
scaling of the delta function is changed to $s^{-1}$, and the 
four-fermion interaction is marginal, $S_{4f}\sim s^0$. Whether or 
not the four-fermion interaction qualitatively changes the theory
of non-interacting quasi-particles described by Eq.~(\ref{seff_fs})
depends on quantum corrections, which can change the scaling 
from marginal to marginally relevant [$S_{4f}\sim \log(s)$] 
or irrelevant [$S_{4f}\sim \log(s)^{-1}$].

The one-loop corrections to the four-fermion interaction
are given by
\beq
\label{bcs_zs}
\delta S_{BCS} \;\sim
\raisebox{-0.1in}{
\resizebox{0.5in}{!}{%
\includegraphics{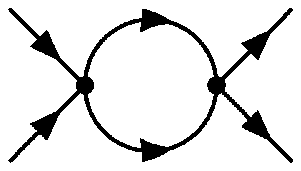}}}, \hspace{1cm}
\delta S_{ZS} \;\sim
\raisebox{-0.22in}{
\resizebox{0.5in}{!}{%
\includegraphics{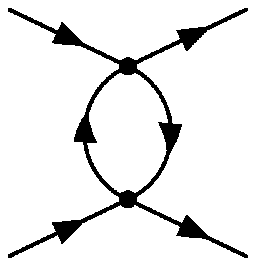}}},  \hspace{1cm}
\delta S_{ZS'} \;\sim
\raisebox{-0.22in}{
\resizebox{0.5in}{!}{%
\includegraphics{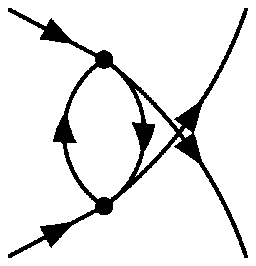}}} .
\eeq
There are two possible scenarios. One possibility is that
the interaction in the BCS channel ($\vec{k}_1=-\vec{k}_2$)
is attractive in some partial wave. In this case the first diagram
in Eq.~(\ref{bcs_zs}) leads to a logarithmic growth of the
interaction. We can illustrate this effect using the $s$-wave
four-fermion interaction defined in Eq.~(\ref{eq:Lagrangian}). For
$\vec{p}_1= -\vec{p}_2$ and $E_1=E_2=E$, the one-loop correction to $C_0$
is given by
\beq 
\label{cor_bcs}
- C_0^2\left(\frac{\kf m}{2\pi^2}\right)
 \log\left(\frac{E_0}{E}\right) \;,
\eeq
where $E_0$ is an ultraviolet cutoff. This result can be interpreted 
as an effective energy-dependent coupling. The coupling constant 
satisfies the renormalization group equation 
\beq 
\label{rge_bcs}
 E\frac{dC_0}{dE} = C_0^2 \left(\frac{\kf m}{2\pi^2}\right)
\hspace{0.25cm}\Rightarrow\hspace{0.25cm}
C_0(E) =\frac{C_0(E_0)}{1+NC_0(E_0)\log(E_0/E)} \;,
\eeq
where $N=\kf m/2\pi^2$ is the density of states. Equation~(\ref{rge_bcs}) 
shows that if the initial coupling is repulsive, $C_0(E_0)>0$, then the 
renormalization group evolution will drive the effective coupling to 
zero. If, on the other hand, the initial coupling is attractive, $C_0(E_0)
<0$, then the effective coupling grows and reaches a pole (called a ``Landau 
pole'') at $E_{\it crit} \sim E_0 \,\exp(-1/(N|C_0(E_0)))$. At the Landau 
pole the effective theory defined by Eq.~(\ref{seff_fs},\ref{sfl_4f})
has to break down. The renormalization group equation does not determine 
what happens at this point, but it is natural to assume that the strong 
attractive interaction leads to the formation of a fermion pair condensate 
in the BCS channel $\langle \psi(\vec{p})\psi(-\vec{p})\rangle$. The 
magnitude of the  difermion condensate as well as the corresponding gap 
in the energy spectrum is easiest to compute if the microscopic interaction 
is weak (if $k_fR<1$). Employing standard methods we can derive the gap 
equation
\beq
\label{4f_gap}
1 = \frac{|C_0|}{2}\int\!\frac{d^3p}{(2\pi)^3} 
 \frac{1}{\sqrt{(E_p-\Ef )^2+\Delta^2}} \;.
\eeq
The infrared divergence in the BCS channel is regulated by the energy 
gap $\Delta$. The gap equation also has a logarithmic ultraviolet
divergence. This divergence can be treated consistently with the 
relation between $C_0$ and $a_0$ derived in Section~\ref{subsec:prototype} 
by using dimensional regularization \cite{Papenbrock:1998wb,Marini:1998}. 
The result is
\beq
\label{gap_bcs}
\Delta = \frac{8\Ef }{e^2}\exp\left(-\frac{\pi}{2\kf |a_0|}\right) \;.
\eeq
The term in the exponent represents the leading term in an expansion 
in $\kf |a_0|$. This means that in order to determine the pre-exponent 
in Eq.~(\ref{gap_bcs}) we have to solve the gap equation at next-to-leading 
order. This correction corresponds to keeping the ZS (``zero sound'')
diagram in Eq.~(\ref{bcs_zs}). In nuclear physics this term is known 
as the ``induced interaction'' \cite{Wambach:1992ik}. In the case 
of a zero-range potential the induced interaction was first computed by 
Gorkov and Melik-Barkhudarov \cite{Gorkov:1961}. It leads to a suppression 
of the $s$-wave gap by a factor $(4e)^{1/3}\simeq 2.2$ 

 For nuclear matter the result given in Eq.~(\ref{gap_bcs}) is not 
very useful, both because the scattering length is large, and because 
effective range corrections are not negligible. We will discuss the 
pairing gap in the limit $a_0\to\infty$ in Section~\ref{sec:unitary}
below. Range corrections in the case of a normal scattering length
were studied in \cite{Papenbrock:1998wb}. A rough estimate of the 
gap at moderate densities can be obtained by replacing $1/(\kf a)$ 
with $\cot[\delta_0(\kf )]$, where $\delta_0(k)$ is the $s$-wave phase 
shift. This estimate gives neutron gaps on the order of 1 MeV at 
nuclear matter density. 
 
  The second scenario arises if the interaction in the BCS channel 
is either repulsive or very weak. In this case the forward scattering 
amplitudes are important.  The interaction is
\beq
\left. U(\hat{p}_4,\hat{p}_3,\hat{p}_2,\hat{p}_1) 
\right|_{\hat{p}_1\cdot\hat{p}_2=\hat{p}_3\cdot\hat{p}_4}
= F(\hat{p}_1\cdot\hat{p}_2,\phi_{12,34}) \;,
\eeq
where $\phi_{12,34}$ is the angle between the plane spanned by 
$\vec{p}_{1,2}$ and $\vec{p}_{3,4}$. The function $F(x,0)$ is 
called the Landau function and its Legendre coefficients are referred 
to as Landau parameters. If spin-dependence is included there is a second set
of Landau parameters commonly denoted $F'_l$.
The Landau parameters remain marginal at one-loop order. 

 The effective field theory characterized by $\vf $ and $F_l$ is 
called Landau Fermi liquid theory \cite{Pines:1966,Baym:1991}. 
The Landau parameters can be related to the compressibility, the 
velocity of zero and first sound, transport coefficients, etc. The 
compressibility of nuclear matter, for example, is given by 
\beq 
\frac{dP}{d\rho} = \frac{\kf ^2}{m^2}
 \frac{1+F_0}{3+F_1}  \;.
\eeq
The coefficients $F_i$ can be extracted from experiment, but 
ultimately we would like to find a systematic method for computing 
the Landau parameters from the underlying nucleon-nucleon interaction. 
One possibility is to use the renormalization group (RG) to integrate
out modes far away from the Fermi surface. A difficulty with this
strategy is the problem of finding suitable initial conditions
for the RG flow. Brown, Friman, and Schwenk proposed to use a free 
space RG to generate a universal low momentum effective interaction 
$\vlowk$ (which we shall discuss in more detail in Section~\ref{sec:wf} 
below). This interaction, evolved to a scale $\Lambda\sim  2\kf $, can 
be used as a starting point for the determination of the Landau 
parameters~\cite{Schwenk:2002fq}.

\subsection{Unnatural Scattering Length}
\label{sec:unitary}

 An important aspect of nuclear physics is the fact that the 
nucleon scattering lengths are anomalously large. The neutron-proton scattering 
length in the $^1S_0$ channel is $-23.71\,{\rm fm}$, and the binding 
energy in the $^3S_1$ (deuteron) channel is $2.2$ MeV. This 
implies that expanding the scattering amplitudes in powers of the 
momentum [as in Eq.~(\ref{eq:scatt})] is not useful, and that powers
of $a_0 k$ have to be kept to all orders. Keeping the first two 
terms in the effective range expansion, the scattering amplitude
can be written as 
\beq
f_0(k) \sim \frac{1}{-1/a_0+r_0k^2/2-ik}
     =  \frac{1}{-1/a_0-ik} 
      \left\{ 1+ \frac{r_0/2}{-1/a_0-ik} + \ldots \right\}\, . 
\eeq
This expansion can be reproduced by keeping the $s$-wave contact 
interaction proportional to $C_0$ to all orders, and treating 
$C_{2i}$ ($i>0$) perturbatively as before. This procedure gives
the correct result, but in dimensional regularization (with 
minimal subtraction) or cutoff regularization the power counting 
of individual diagrams is not manifest. This is easily seen in 
dimensional regularization where $C_0\to \infty$ as $a_0\to\infty$. 
As a consequence, individual diagrams diverge in the limit of 
a large scattering length even though the sum of all diagrams
is finite. Kaplan, Savage, and Wise proposed a modified version
of dimensional regularization (power divergence subtraction, PDS)
in which poles in lower dimensions are subtracted, and power
counting is manifest \cite{Kaplan:1998we}.

 Interest in many-body systems with a large two-particle 
scattering length arises not only in nuclear physics, but 
also in atomic physics. It is now possible to create cold atomic
gases in which the scattering length $a_0$ of the atoms can be 
adjusted experimentally using Feshbach resonances, see \cite{Regal:2005} 
for a review. If the density is low the atoms can be described
as pointlike nonrelativistic particles that carry a ``spin'' 
label which characterizes the hyperfine quantum numbers of the 
atoms. A Feshbach resonance arises if a molecular bound state in 
a closed hyperfine channel crosses near the threshold of a lower 
``open'' channel. Because the magnetic moments of the open and 
closed states are in general different, Feshbach resonances can be 
tuned using an applied magnetic field.  At resonance the two-body 
scattering length in the open channel diverges, and the cross 
section $\sigma$ is limited only by unitarity, $\sigma(k) = 4\pi/k^2$ 
for low momenta $k$. In the unitarity limit, details about the 
microscopic interaction are lost, and the system displays universal 
properties.

 A dilute gas of \emph{any} fermions in the unitarity limit 
is a strongly coupled
quantum liquid that exhibits universality. 
At low density we are interested in the limit 
$\kf a_0\to\infty$ and $\kf r_0\to 0$.  From 
dimensional analysis it is clear that the energy per particle at zero
temperature has to be proportional to energy per particle of a free Fermi 
gas at the same density,
\beq
\frac{E}{A} = \xi \Big(\frac{E}{A}\Big)_0 = \xi 
\frac{3}{5}\Big(\frac{\kf ^2}{2m}\Big) \;.
\eeq
The constant $\xi$ is universal, \textit{i.e.}, independent of the details of 
the system. Similar universal constants govern the magnitude of the 
gap in units of the Fermi energy and the equation of state at finite
temperature. 

 Calculating these universal constants is clearly a very challenging
task -- many-body diagrams containing $C_0$ have to be summed to all
orders. One possibility is to do the calculation numerically, using
diffusion or imaginary time path integral Monte Carlo methods as described
in Section~\ref{sec_unit_latt}. It is also desirable to find systematically 
improvable analytical approaches. Analytical methods offer the possibility
to systematically include higher order terms in the interaction (range
corrections, explicit pions, three-body forces, etc.) and to determine
real time properties that are hard to access numerically. 
A number of analytical methods have been considered, such as an expansion 
in the number of fermion species \cite{Furnstahl:2002gt,Nikolic:2006} or 
the number of spatial dimensions (which is related to the hole-line
expansion of Brueckner, Bethe, and Goldstone)~\cite{Steele:2000qt,Schafer:2005kg}. 
In the following we shall discuss a proposal by Nussinov \& Nussinov
to perform an expansion around $d=4-\epsilon$ spatial dimensions. 
Epsilon expansions are well known in the theory of critical phenomena.
An interesting aspect of the epsilon expansion in nuclear physics is
that both many-body and few-body system can be studied
\cite{Rupak:2006jj}.
 
Nussinov \& Nussinov observed that the fermion many-body system in 
the unitarity limit reduces to a free Fermi gas near $d=2$ spatial 
dimensions, and to a free Bose gas near $d=4$ \cite{Nussinov:2004}. 
Their argument was based on the behavior of the two-body wave function 
as the binding energy goes to zero. For $d=2$ it is well known that the 
limit of zero binding energy corresponds to an arbitrarily weak potential. 
In $d=4$ the two-body wave function at $a_0=\infty$ has a $1/r^2$ behavior 
and the normalization is concentrated near the origin. This suggests 
the many-body system is equivalent to a gas of non-interacting bosons.
A systematic expansion based on the observation of Nussinov \& Nussinov
was studied by Nishida and Son \cite{Nishida:2006br}. In
this section we shall explain their approach. 

We begin by restating the 
argument of Nussinov \& Nussinov in the effective field theory language. 
For simplicity we shall work with dimensional regularization and minimal
subtraction. In this case $a_0\to\infty$ corresponds to $C_0\to\infty$. 
The fermion-fermion scattering amplitude is given by
\beq 
 f(p_0,\vec{p}) =  \left(\frac{4\pi}{m}\right)^{d/2}
 \left[\Gamma\left(1-\frac{d}{2}\right)\right]^{-1} 
 \frac{i}{\left(-p_0+E_p/2-i\delta\right)^{\frac{d}{2}-1}}\; ,
\eeq
where $\delta\to 0+$. As a function of $d$ the Gamma function has poles 
at $d=2,4,\ldots$ and the scattering amplitude vanishes at these points. 
Near $d=2$ the scattering amplitude is energy and momentum independent.
For $d=4-\epsilon$ we find
\beq
\label{A_4-eps}
 f(p_0,\vec{p}) =  \frac{8\pi^2\epsilon}{m^2}
 \frac{i}{p_0-E_p/2+i\delta} + {\cal O}(\epsilon^2) \; .
\eeq
We observe that at leading order in $\epsilon$ the scattering amplitude 
looks like the propagator of a boson with mass $2m$. The boson-fermion
coupling is $g^2=(8\pi^2\epsilon)/m^2$ and vanishes as $\epsilon\to 0$. 
This suggests that we can set up a perturbative expansion involving 
fermions of mass $m$ weakly coupled to bosons of mass $2m$. A difermion
field can be introduced using the Hubbard-Stratonovich trick. The 
difermion self-coupling is proportional to $1/C_0$ and vanishes in 
the unitarity limit. The Lagrangian is 
\beq
\label{L_hs}
{\cal L}= \Psi^\dagger\left[
     i\partial_0+\sigma_3\frac{\vec\nabla^2}{2m}\right]\Psi
  + \mu\Psi^\dagger\sigma_3\Psi
  +\left(\Psi^\dagger\sigma_+\Psi\phi + \mbox{h.c.} \right) \;,\nonumber
\eeq
where $\Psi=(\psi_\uparrow,\psi_\downarrow^\dagger)^T$ is a two-component 
Nambu-Gorkov field, $\sigma_i$ are Pauli matrices acting in the Nambu-Gorkov 
space and $\sigma_\pm=(\sigma_1\pm i\sigma_2)/2$. 

In the superfluid phase
$\phi$ acquires an expectation value. We write 
\beq
\label{phi_exp}
 \phi = \phi_0 + g\varphi \;, \hspace{1cm}
   g  =\frac{\sqrt{8\pi^2\epsilon}}{m}
       \left(\frac{m\phi_0}{2\pi}\right)^{\epsilon/4} \;,
\eeq
where $\phi_0=\langle\phi\rangle$. The scale $M^2=m\phi_0/(2\pi)$ was
introduced in order to have a correctly normalized boson field. The scale 
parameter is arbitrary, but this particular choice simplifies some of the 
algebra. In order to get a well-defined perturbative expansion we add and 
subtract a kinetic term for the boson field to the Lagrangian. We include 
the kinetic term in the free part of the Lagrangian
\beq
\label{L_eps_0}
{\cal L}_0 = \Psi^\dagger\left[i\partial_0+\sigma_3\frac{\vec\nabla^2}{2m}
     + \phi_0(\sigma_{+} +\sigma_{-})\right]\Psi
     + \varphi^\dagger\left(i\partial_0
        + \frac{\vec\nabla^2}{4m}\right)\varphi \;,
\eeq
and the interacting part is 
\beq
\label{L_eps_I}
{\cal L}_I = g\left(\Psi^\dagger\sigma_+\Psi\varphi + \mbox{h.c.}\right)
     + \mu\Psi^\dagger\sigma_3\Psi 
     - \varphi^\dagger\left(i\partial_0
        + \frac{\vec\nabla^2}{4m}\right)\varphi \;.
\eeq
Note that the interacting part generates self-energy corrections to 
the boson propagator which, by virtue of Eq.~(\ref{A_4-eps}), cancel 
against the kinetic term of boson field. We have also included the 
chemical potential term in ${\cal L}_I$. This is motivated by the fact 
that near $d=4$ the system reduces to a non-interacting Bose gas and
$\mu\to 0$. We will count $\mu$ as a quantity of ${\cal O}(\epsilon)$. 

 The Feynman rules are quite simple. The fermion and boson propagators 
are
\beq
  \label{eps_prop}
  G(p_0,\vec p) = \frac{i}{p_0^2-E_p^2-\phi_0^2}
  \left[\begin{array}{cc}
      p_0+E_p &  -\phi_0\\
      -\phi_0        & p_0-E_p
  \end{array}\right]  \ , \hspace{0.5cm}
  D(p_0, \vec p) = \frac{i}{p_0-E_p/2} \;, 
\eeq
and the fermion-boson vertices are $ig\sigma_\pm$. Insertions of the
chemical potential are $i\mu\sigma_3$. Both $g^2$ and $\mu$ are
corrections of order $\epsilon$. There are a finite number of
one-loop diagrams that generate $1/\epsilon$ terms. All other
diagrams are finite, and the $\epsilon$ expansion is well defined.

 The ground-state energy is determined by diagrams with no external 
legs. The first diagram is the free fermion loop which is ${\cal O}(1)$ in 
the epsilon expansion. We get
\beq
\raisebox{-0.21in}{ \resizebox{0.5in}{!}{%
\includegraphics{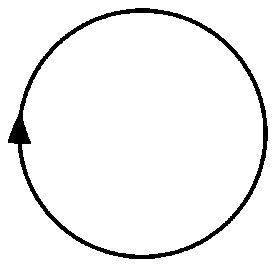}}} \;\;
     =-\int\!\frac{d^dp}{(2\pi)^d}\, \sqrt{E^2_{p}+\phi_0^2}
     =  \frac{\phi_0}{3}\left[ 
   1 + \frac{7-3(\gamma+\log(2))}{6}\, \epsilon \right]
  \left(\frac{m\phi_0}{2\pi}\right)^{d/2} \;.
\eeq
An insertion of $\mu$ is also ${\cal O}(1)$ because the loop diagram is 
divergent in $d=4$. We find
\beq
\raisebox{-0.21in}{ \resizebox{0.5in}{!}{%
\includegraphics{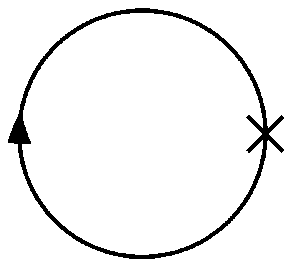}}} \;\;
   = \mu\int\!\frac{d^dp}{(2\pi)^d}
     \frac{E_p}{\sqrt{E^2_{p}+\phi_0^2}} 
   = -\frac{\mu}{\epsilon}\left[ 
   1 + \frac{1-2(\gamma-\log(2))}{4}\, \epsilon \right]
  \left(\frac{m\phi_0}{2\pi}\right)^{d/2} \;.
\eeq
Graphs with extra insertions of $\mu$ follow the naive epsilon counting 
and are at least ${\cal O}(\epsilon^2)$. Nishida and Son also computed the leading 
two-loop contribution which is ${\cal O}(\epsilon)$ because of the factor of $g^2$ 
from the vertices. The result is 
\beq
 \raisebox{-0.21in}{ \resizebox{0.5in}{!}{%
\includegraphics{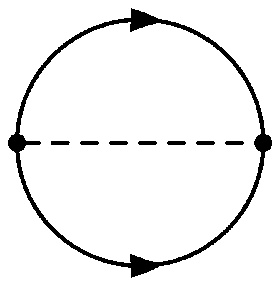}}} \;\;
   = -C\epsilon  \left(\frac{m\phi_0}{2\pi}\right)^{d/2} \;,
\eeq
where the dashed line denotes the difermion propagator, and $C\simeq 
0.14424$. 

We can now determine the minimum of the effective potential. 
We find $\phi_0 = (2\mu)/\epsilon( 1 + C'\epsilon + {\cal O}(\epsilon^2))$
with $C'=3C-1+\log(2)$. The value of the effective potential at $\phi_0$ 
determines the pressure and $n=\partial P/\partial \mu$ gives the density. 
From the density we can compute the Fermi momentum ($n\sim \kf ^d$ in $d$ 
dimensions), and the relationship between the Fermi energy $\ef =
\kf ^2/2m$ and $\mu$ determines the universal parameter $\xi=\mu/
\ef $. We find 
\beq 
\xi = \frac{1}{2}\epsilon^{3/2} 
      + \frac{1}{16}\epsilon^{5/2}\log(\epsilon)
      - 0.025\epsilon^{5/2} + \ldots 
    = 0.475 \hspace{0.25cm} (\epsilon=1) \;,
\eeq
which agrees quite well with the result of fixed-node quantum 
Monte Carlo calculations. The calculation was extended to ${\cal O}(
\epsilon^{7/2})$ by Arnold et al.~\cite{Arnold:2006fr}. Unfortunately,
the next term is very large and it appears necessary to combine the 
expansion in $4-\epsilon$ dimensions with a $2+\epsilon$ expansion 
in order to extract useful results. The $\epsilon$ expansion has also 
been applied to the calculation of the gap \cite{Nishida:2006br}. At 
next-to-leading order the result is $\Delta = 0.62 \ef $. 
Somewhat surprisingly, this result is quite close to the naive 
$a_0\to\infty$ limit of the BCS result Eq.~(\ref{gap_bcs}), provided
the induced interaction term is taken into account.


\section{EFT FOR FINITE NUCLEI AND NUCLEAR MATTER}
  \label{sec:nuclei}
  
In this section, we survey the wide range of pioneering applications
of EFT to nonrelativistic finite density nuclear systems. 
These frontiers are rapidly evolving and most results are immature, 
so we focus on general illustrative aspects.

\subsection{Pion Physics From Chiral EFT}

To apply EFT to finite nuclei and nuclear matter, we must first
consider the appropriate degrees of freedom.  
Applications to sufficiently low-density
systems such as dilute neutron matter are possible with nucleons only.
These are called ``pionless'' effective field theories.  
In such an EFT, the pion is a heavy degree of freedom whose effects
are mimicked by contact terms.
This EFT breaks down when external momenta are comparable to the
pion mass, so that pion exchange is resolved.
This does not automatically translate into a clear limit on its applicability
to finite nuclei; pionless EFT is successful for at least the ground
states
of the deuteron and triton and its limits for heavier nuclei are not
yet known~\cite{Bedaque:2002mn}. 

However, given that the Fermi momentum $\kf$ for the interior
of heavy nuclei is about twice the pion mass, one expects the
pion must be treated as a long-range degree-of-freedom in a free-space EFT
applicable to most nuclei. 
Chiral effective field theories for nucleons 
incorporate the pion systematically as
the (near) Goldstone boson of approximate and
spontaneously broken chiral symmetry,
expanding about the massless pion limit.
The functional dependence on the QCD quark masses is captured in
perturbation theory and the dependence on the strong coupling is
contained in universal parameters to be determined from data or
direct numerical calculations of QCD.
Chiral EFT in nuclear physics
originated with the seminal work of Weinberg and van Kolck and
collaborators in the early
1990's~\cite{Weinberg:1990rz,Weinberg:1991um,Ordonez:1992xp,Weinberg:1992yk,%
Ordonez:1993tn,VanKolck:1994yi,Ordonez:1995rz},
and there has been active development ever 
since~\cite{Beane:2000fx,Bedaque:2002mn,Epelbaum:2005pn}.

The most commonly applied chiral EFT Lagrangians at present have
nonrelativistic nucleons and pions as degrees of freedom
based on the ``heavy-baryon'' formalism, which eliminates anti-nucleons
and organizes relativistic corrections~\cite{Bedaque:2002mn}.
As usual, renormalization can be carried out because all interactions
consistent with QCD symmetries are included, which allows regulator
dependence to be absorbed.
To organize the EFT in a systematic hierarchy we need a power counting
but the optimal scheme is not yet settled.  Both practical and formal
questions are being argued  and different schemes are under 
investigation~\cite{Beane:2001bc,Nogga:2005hy,Birse:2005um,Epelbaum:2006pt}.
In all cases,
chiral symmetry dictates that pion interactions are accompanied
by derivatives (because they are Goldstone
bosons) or powers of the pion mass (Ward identity constraints from QCD), 
which then yields ratios of characteristic momenta and
$m_\pi$ to the scale of excluded physics, such as heavier meson
exchange, as expansion parameters.
Relativistic corrections are organized in powers momenta over the
nucleon mass.

For applications to nuclear structure,
an energy-independent nucleons-only 
potential is desirable (and required for many of the methods discussed
below); it can be derived from the chiral Lagrangian by a unitary 
transformation
method that decouples the nucleons from \emph{explicit} pion
fields, leaving static pion-exchange interactions 
and regulated contact terms~\cite{Epelbaum:2005pn}. 
At present these potentials are organized by
a power counting proposed by Weinberg and
then iterated with a Lippmann-Schwinger equation for two-body scattering
or other nonperturbative methods for bound-state properties
with more nucleons.
A momentum-space cutoff is used
for technical reasons, which means
the advantages of dimensional regularization 
we saw for short-range interactions at finite density are not available.

For Weinberg power counting there is a formula 
analogous to Eq.~(\ref{eq:naturalpc})
that identifies the order in the EFT expansion at which a given term
in the potential contributes.
This yields a hierarchy of terms with
increasing derivatives and pion exchanges and,
perhaps most important for finite density applications to be tractable, 
a hierarchy of many-body forces.
At leading order (LO), there is one-pion exchange and two no-derivative contact
terms.  The next-to-leading-order (NLO) adds the first two-pion
exchange contributions, which are important for the mid-range nuclear attraction.
At present, NN interactions go to up to N$^3$LO, which includes 24 
constants for the contact terms
(not including isospin violation) that are determined by
fits to NN scattering.
The best fits have a $\chi^2$/dof comparable to the best 
phenomenological potentials~\cite{Entem:2003ft,Epelbaum:2004fk}. 

\begin{figure}[t!]
\centering
  \includegraphics*[width=2.2in]{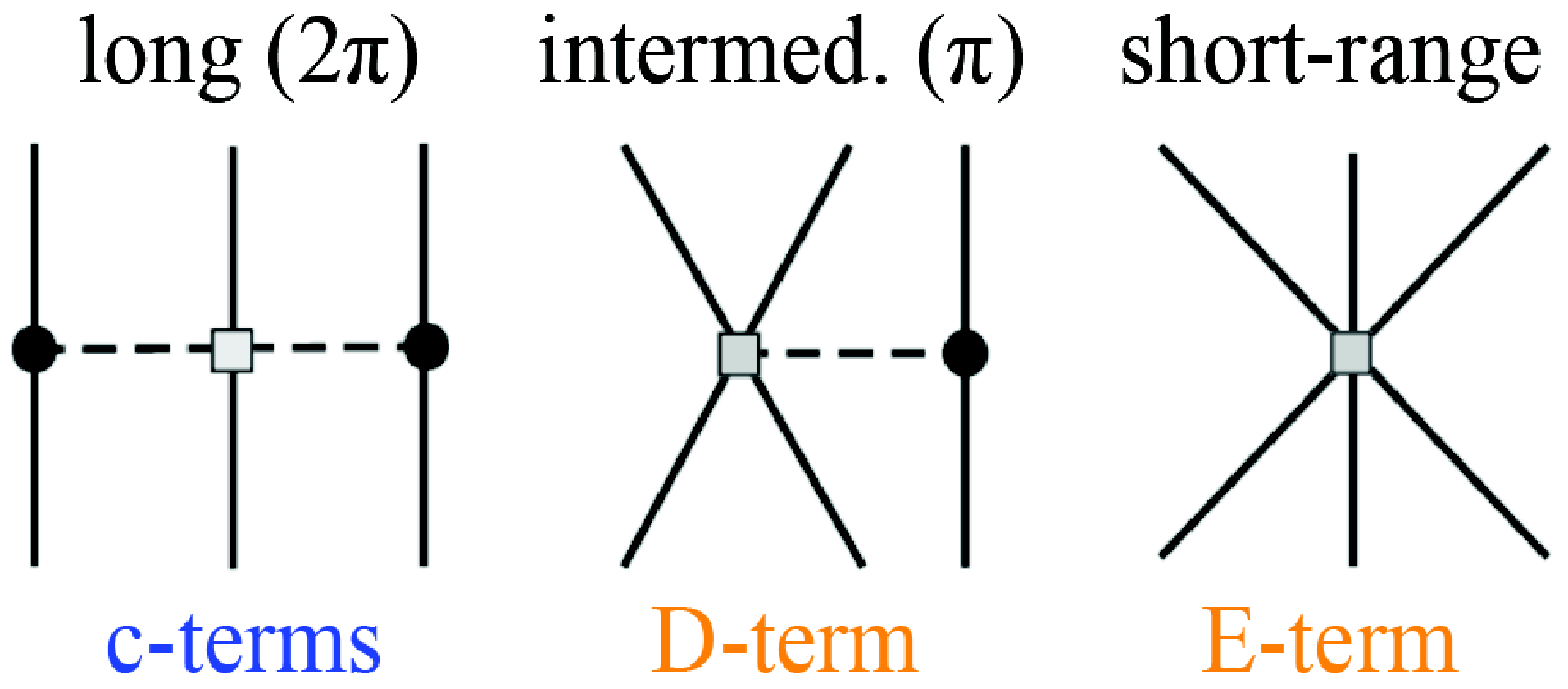}
    \hspace*{.5in}       
  \includegraphics*[width=0.7in]{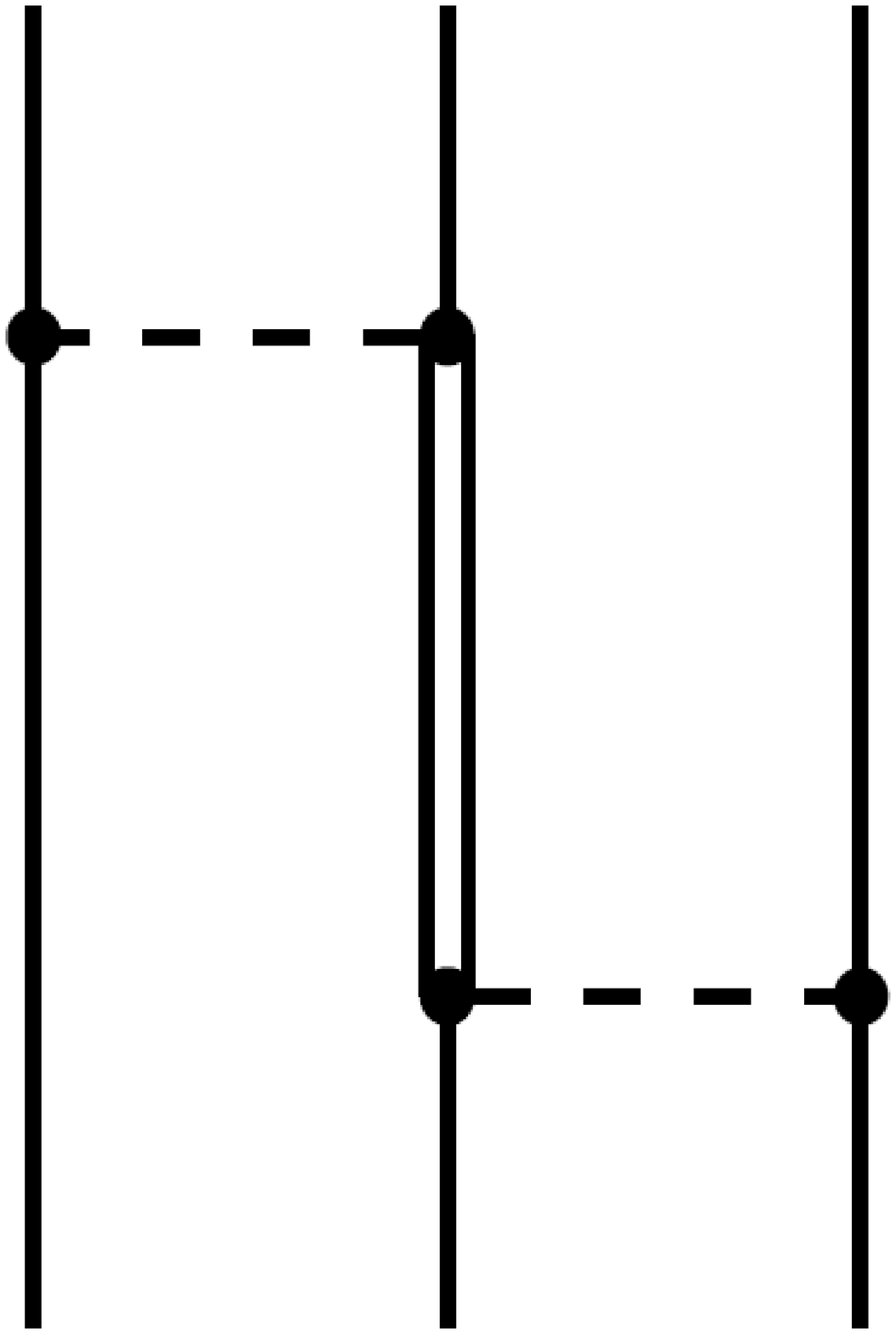}
 \caption{Leading three-body contributions in chiral EFT.
 Left: N$^2$LO terms in a EFT without $\Delta$'s (dashed line is the pion).
 Right: NLO contribution with explicit $\Delta$'s (double line).}
 \label{fig:3NF}
\end{figure}

Three-nucleon forces (3NF) appear first at N$^2$LO and are shown
on the left in Fig.~\ref{fig:3NF}.   
There are parameters associated with long-range two-pion exchange
(four constants fit to $\pi N$ or $NN$ scattering), 
mid-range one-pion exchange (one constant), and purely
short-range (one constant) parts.
The extension to N$^3$LO is in progress and involves
many additional diagrams but no additional parameters.
However, there are sizable
uncertainties at present in determining the long-range 3NF 
parameters from $\pi$N or NN scattering, which
translates into significant uncertainties at finite density.
The 4N interaction appears first at N$^3$LO in the form of
long-range pion exchange and is parameter free~\cite{Epelbaum:2007us}.
The quantitative suppression of many-nucleon forces
predicted by chiral power counting is consistent with
binding-energy calculations in light 
nuclei~\cite{Epelbaum:2005pn,Rozpedzik:2006yi}, 
but much remains to be tested
in larger systems.

Even after we have specified a power counting and the order in the expansion,
there is not a unique EFT potential because one can choose different cutoffs. 
Calculations of observables should be independent of the cutoff at
the level of the truncation error determined by the missing orders.
By comparing calculations with varied cutoffs one can test whether
the EFT is working and put a bound on the theoretical error.
The precision EFT potentials currently available for nuclear structure
have cutoffs in a rather narrow range close to the expected breakdown
scale of the EFT, about 450--600\,MeV (\textit{cf.}\ the $\rho$ or $\omega$ meson
mass), which is consistent with the prescription of
Lepage~\cite{Lepage:1989hf,Lepage:1997cs}.  
In practice, lower cutoffs would mean large truncation errors
(\textit{i.e.}, the expansion parameter $q/\Lambda_c$ gets too small)
while larger cutoffs create implementation
problems with increasingly singular
(at short distances) potentials from multiple pion exchange.
Within this cutoff range there is no penalty for iterating sub-leading
potential terms, which violates some power countings, because
the truncation and iteration errors are the same size~\cite{Bedaque:2002mn}.

Recent surveys of on-going applications of chiral potentials
to scattering and to properties of few-body
nuclei can be found in Refs.~\cite{Bedaque:2002mn,Epelbaum:2005pn}.
Among the developments most relevant to finite density is
work to add the 
$\Delta(1232)$-isobar resonance explicitly to the chiral EFT Lagrangian; 
this was part of the original explorations by
van Kolck \emph{et al.},
but has only recently been reconsidered for energy-independent 
potentials~\cite{Krebs:2007rh,Epelbaum:2007sq}.
The $\Delta$ is considered important because of its low-excitation
energy (the mass difference to the nucleon is about 300\,MeV) and its
strong coupling to the $\pi N$ system.
Including it would resum important contributions and improve the pattern
of convergence.
In this scheme, the leading 3NF term comes from pion exchange with an intermediate
$\Delta$ (right diagram in Fig.~\ref{fig:3NF}) and appears at NLO.
As this and other developments mature, in parallel there will be
applications to finite nuclei.
Indeed, since the energy-independent potentials
take the same form as phenomenological
non-local potentials, almost all 
conventional few- and many-body methods are immediately available.

\subsection{Wave Function Methods}
  \label{sec:wf}

There are a wide variety of methods available to determine properties
of few-body systems given an internucleon potential.  These all in some way
involve solving for the approximate wave function of the system.
If we arbitrarily set the cross-over from few-body to many-body nuclei at
$A=8$, 
the choice of methods dwindles to a few: Green's function
Monte Carlo (GFMC), no-core shell model (NCSM), and coupled
cluster (CC).
The GFMC approach~\cite{Pieper:2004qh,Pieper:2007ax} 
has had great success up to $A=12$ (and extensions
using auxiliary field methods promise to go much further), but is
limited at present to local potentials, 
\textit{i.e.}, diagonal in coordinate representation,
which excludes current chiral EFT interactions.
However, both NCSM and CC methods are compatible with energy-independent
chiral potentials including many-body
forces~\cite{Nogga:2005hp,Hagen:2007hi}.

The NCSM diagonalizes the Hamiltonian in a harmonic oscillator basis
with all nucleons active (hence ``no core'').
Lanczos methods allow the extraction of the lowest eigenvalues and
eigenvectors from spaces up to dimension $10^9$ (and growing with computer
hardware and software advances),
but the matrix size grows rapidly with $A$ and the maximum oscillator
excitation energy $N_{\rm max}\hbar\Omega$.
For a given $A$, the convergence of observables with 
$N_{\rm max}$ depends strongly on
the nature of the potential. 
Chiral EFT Hamiltonians are softer than conventional
nuclear potentials (\textit{i.e.}, smaller high-momentum contributions,
which means less coupling to high oscillator states), but adequate convergence
with 3NF still requires too large a basis beyond the lightest nuclei.  
Therefore Lee-Suzuki 
transformations of the potential, which are unitary order-by-order
in a cluster expansion, are applied
to decouple included and excluded oscillator states, 
greatly reducing the size of the model space needed.
This procedure has many demonstrated 
successes~\cite{Nogga:2005hp,Navratil:2007we} 
although there are drawbacks, such 
as distortions of  long-range physics, problems with
extrapolations of energies, 
and the loss of the variational principle~\cite{Stetcu:2006zn}.

\begin{figure}[t!]
\centering
      \includegraphics*[width=6.2in]{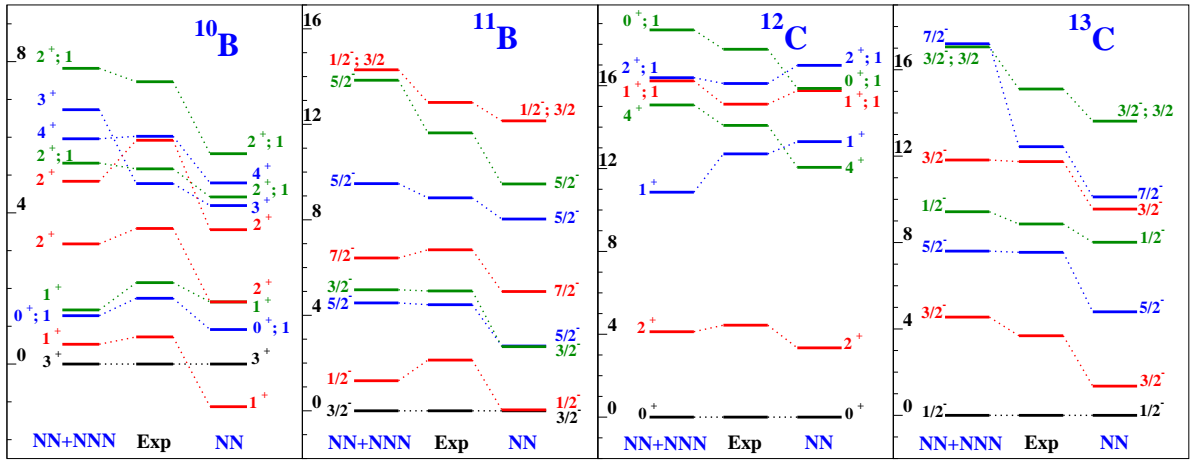}
 \caption{Excitation energies (in MeV) 
 of selected levels in four $p$--shell
 nuclei~\cite{Navratil:2007we} calculated using the N$^3$LO potential
 of Ref.~\cite{Entem:2003ft}. Calculations with NN-only and with N$^3$LO NN plus N$^2$LO
 3NF are compared to experiment.}
 \label{fig:n3lonuclei}       
\end{figure}

Recent state-of-the-art NCSM calculations of excitation energies
for four $p$--shell nuclei are shown in
Fig.~\ref{fig:n3lonuclei} for a single N$^3$LO potential with and without
the N$^2$LO 3NF~\cite{Navratil:2007we}.
(The mismatch in orders means that this 
calculation is not yet completely consistent from the EFT perspective.)
It is evident that the fine structure of the nuclear spectra
is uniformly improved with the three-body contribution.
Of particular note is the ground state of $^{10}$B
and the splittings of spin-orbit partners throughout.
The sensitivity of three-body parameters to particular
observables (\textit{e.g.}, $^6$Li quadrupole moment, the lowest $1^+$ states in
$^{10}$B) suggests that fits of 3NF parameters will improve with input
from more than $A=3,4$ systems~\cite{Navratil:2007we}.

How do we tell if an EFT-based interaction
used by a wave function method is working as advertised?
One way is to do comparative calculations at different orders in
the EFT with a range of cutoffs.
Since cutoff variation is absorbed by the contact interactions, which
scale as powers of the inverse cutoff,
the relative variation of the potential energies 
over this range should decrease systematically
according to the power of omitted contact interactions
(and assuming a typical momentum
scale $\approx 130\,$MeV).
Binding energy variations will be larger because of cancellations
in nuclear systems, which amplify the role of higher orders
(including many-body forces).
Nogga has shown that such estimates are consistent with calculations
in $^3$H~\cite{Nogga:2006ir}.
For $^4$He, he concludes
that the EFT estimate of 2\% for the ratio of 3NF to NN potential
energies is consistent with observed ratios of roughly 5\%~\cite{Nogga:2006ir} 
and
preliminary calculations of the 4NF contribution were found to be
as small as expected~\cite{Rozpedzik:2006yi}.
All of these tests will need to be repeated for larger nuclei 
as reliable calculations
become possible.

Renormalization group (RG) methods applied in free space to chiral EFT 
interactions are a promising avenue to calculating larger nuclei.
These methods prescribe how each matrix element of the potential
(and other operators) in a discretized momentum basis
must evolve under changes in the ``resolution scale'' so that
observables are unchanged.  
(Since the potential is not an observable, we are always free to make
unitary transformations.)
The resolution scale is changed by lowering a cutoff in relative momentum
(``$\vlowk$''~\cite{Bogner:2003wn}) 
or using a flow equation for the Hamiltonian
(``similarity RG''~\cite{Bogner:2006pc}) or by tailored unitary
transformations (``UCOM''~\cite{Roth:2005pd}). 
The result is a decoupling of high- and low-momentum dependence
without modifying long-distance interactions,
leading to low-momentum potentials that are more perturbative,
such that convergence in harmonic oscillator
bases is dramatically accelerated~\cite{Bogner:2007rx}.
Such potentials can be applied without Lee-Suzuki transformations
in the NCSM and maintain the variational principle. 
Since the transformations are unitary, the EFT truncation error is
unchanged, in contrast to the RG evolution of a chiral EFT at fixed order
to low cutoff.  
However, the evolution of the NN potential is inevitably accompanied by
the evolution of the three- and higher-body potentials. 
The
latter is not yet implemented but is instead approximated by
fitting the N$^2$LO chiral interaction at each
cutoff~\cite{Nogga:2004ab}, which introduces a theoretical error.

These low-momentum potentials show great promise for the CC method,
which has been highly developed in \textit{ab initio} quantum chemistry
but only recently revived for nuclear applications, including the 
development of CC theory for three-body
Hamiltonians~\cite{Hagen:2007ew,Hagen:2007hi}. 
Coupled cluster calculations are based on a potent exponential ansatz for
the ground-state wave function, $|\psi\rangle = e^{\widehat T}|\phi\rangle$,
where $|\phi\rangle$ is a simple reference state, typically a harmonic
oscillator Slater determinant.  The operator $\widehat T$ is specified by
amplitudes for a truncated sum of operators creating one-particle--one-hole,
two-particle--two-hole, \textit{etc.\ } excitations.  The amplitudes are found
from nonlinear equations whose solution scales very gently with the size of the
nucleus and model space. 

As with the NCSM, convergence is accelerated with low-momentum potentials
and particularly promising is the calculation of 3NF contributions, 
which are the
most expensive component.
The 3NF potential is rewritten in terms of normal-ordered creation and
destruction operators with respect to $|\phi\rangle$ (instead of the vacuum),
which recasts the 3NF into an expectation value in $|\phi\rangle$, 
one- and two-body pieces, and the remaining 3NF part.
In the hierarchy of contributions to a CC calculation,
only the last piece is expensive to calculate, but
recent CC calculations of $^4$He found it to be
negligible~\cite{Hagen:2007hi}. 
If this results persists for larger nuclei, calculations of $A=100$
or beyond will be feasible in the near future!  The present limit for NCSM
is much lower, around $A=16$
but could be extended using
importance sampling methods that pick out the most important 
basis states~\cite{Roth:2007sv}, 
if they can be implemented in a size-extensive way.

The NCSM and CC wave function methods
apply EFT (and RG) only to create the input potential
and not in solving the many-body problem.
There is also the possibility of a more EFT-like treatment, such as
the pioneering work to apply EFT
to the shell model by Stetcu, Barrett and van Kolck~\cite{Stetcu:2006ey}.
(See Ref.~\cite{Haxton:2007hx} for 
a completely different application of EFT methods to the shell model.)
These authors formulate an EFT in the harmonic oscillator 
basis, where the restricted model space generates all interactions consistent
with the underlying symmetries.
The parameters are directly determined in the model space rather than 
fitting in free space and transforming the interaction.
The oscillator frequency sets an infrared cutoff 
$\lambda \sim \sqrt{M_N\hbar\Omega}$ while the ultraviolet cutoff
is $\Lambda \sim \sqrt{M_N(N_{\rm max}+3/2)\hbar\Omega}$.
Within each model space, a set of observables is used to fix the EFT parameters
and then other observables are calculated.  
The EFT works if cutoff dependence
decreases with decreasing $\lambda$ and increasing $\Lambda$; in that case an
extrapolation to the continuum limit $\hbar\Omega\rightarrow 0$ with
$N_{\rm max}\rightarrow 0$ with $\Lambda$ fixed is made.  At the end, one
takes $\Lambda\rightarrow \infty$.  The first application with a pionless
theory up to $A=6$ is encouraging and motivates generalizations
to the pionful theory and to other many-body methods~\cite{Stetcu:2006ey}.

\subsection{EFT on the Lattice}
\label{sec_unit_latt}

 We have seen that chiral effective field theory potentials have been 
used successfully in connection with standard numerical many-body 
approaches such as coupled cluster or the no-core shell 
model. A disadvantage of these methods is that they rely on the 
existence of a potential, which is not an observable, and as a 
consequence scheme and renormalization scale invariance are not 
manifest. A numerical few and many-body method that is based
directly on the effective Lagrangian is the euclidean lattice 
path integral Monte Carlo method. Euclidean lattice calculations
are standard in the context of QCD but, except for some isolated 
attempts \cite{Brockmann:1992in,Muller:1999cp}, have been introduced 
in nuclear physics only 
recently~\cite{Lee:2004qd,Lee:2005it,Seki:2005ns}.

 In the following we shall introduce the method in the case of 
a simple $s$-wave contact interaction ${\cal L}= -C_0(\psi^\dagger 
\psi)^2/2$. More sophisticated interactions involving higher partial 
wave terms and explicit pions are discussed in \cite{Borasoy:2007vi}.
The usual strategy for dealing with the four-fermion interaction is 
to use a Hubbard-Stratonovich transformation. The partition function 
can be written as \cite{Lee:2004qd}
\beq
\label{z_latt}
Z = \int\! Ds\, Dc\, Dc^{\ast} \exp\left[-S\right]  \;,
\eeq
where $s$ is the Hubbard-Stratonovich field and $c$ is a Grassmann field. 
$S$ is a discretized euclidean action
\beqa
\label{s_latt}
S  &=& 
 \sum_{\vec{n},i}\left[  e^{-\hat\mu\alpha_{t}}c_{i}^{\ast}
   (\vec{n})c_{i} (\vec{n}+\hat{{\bf 0}})-e^{\sqrt{-C_0\alpha_{t}}
  s(\vec{n})+\frac{C_0\alpha_{t}}{2}}(1-6h)c_{i}^{\ast}
   (\vec{n})c_{i} (\vec{n})\right] \nonumber\\
& & \hspace{0.3cm}\mbox{} 
   -h\sum_{\vec{n},{\bf l}_{s},i}\left[  
   c_{i}^{\ast}(\vec{n})c_{i}(\vec{n}
   +\hat{{\bf l}}_{s})+c_{i}^{\ast}(\vec{n})c_{i}
  (\vec{n}-\hat{{\bf l}}_{s})\right]  +\frac{1}{2}\sum_{\vec{n}}s^{2}(\vec{n}).
\eeqa
Here $i$ labels spin and $\vec{n}$ labels lattice sites. Spatial and
temporal unit vectors are denoted by $\hat{{\bf l}}_s$ and $\hat{{\bf 0}}$, 
respectively. The temporal and spatial lattice spacings are $b_\tau$
and $b$, and the dimensionless chemical potential is given by $\hat{\mu}
=\mu b_\tau$. We define $\alpha_t$ as the ratio of the temporal and 
spatial lattice spacings and $h=\alpha_t/(2\hat{m})$. The action 
(\ref{s_latt}) is quadratic in the fermion fields, and can be 
simulated using a variety of methods such as determinant or hybrid 
Monte Carlo. Note that for $C_0<0$ the action is real and importance
sampling is possible. 

\begin{figure}[t!]
\centering
\includegraphics[width=3in]{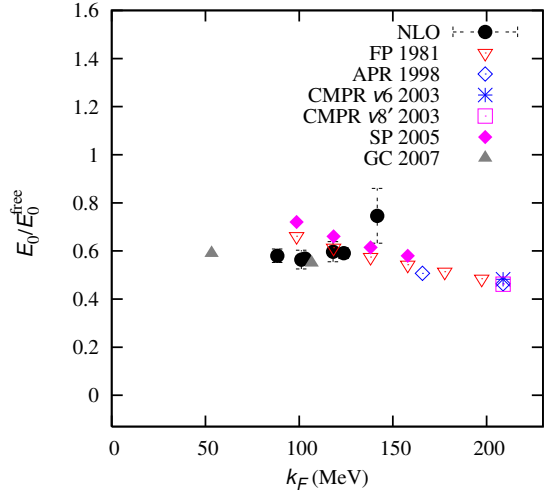}
\caption{
Lattice results for the energy per particle of a dilute Fermi gas 
from Borasoy \textit{et al.}~\cite{Borasoy:2007vi}. We show the energy per particle 
in units of the energy per particle of the free system as a function 
of the Fermi momentum. The solid dots are the lattice results. For 
comparison, we also show results from wave-function-based many-body 
calculations (see~\cite{Borasoy:2007vi}).}
\label{fig_latt}
\end{figure}

 The four-fermion coupling is fixed by computing the sum of all 
particle-particle bubbles as in Section~\ref{sec:unitary} but with the 
elementary loop function regularized on the lattice. Schematically, 
\beq 
\frac{m}{4\pi a_0} = \frac{1}{C_0} 
   + \frac{1}{2} \sum_{\vec{p}}\frac{1}{E_{\vec{p}}} ,
\eeq
where the sum runs over discrete momenta on the lattice and $E_{\vec{p}}$ 
is the lattice dispersion relation. A detailed discussion of the 
lattice regularized scattering amplitude can be found in 
\cite{Chen:2003vy,Beane:2003da,Lee:2004qd}. For a given scattering 
length $a_0$ the four-fermion coupling is a function of the lattice 
spacing. The continuum limit correspond to taking the temporal 
and spatial lattice spacings $b_\tau$, $b$ to zero
\beq 
 b_\tau\mu\to 0, \hspace{1cm} bn^{1/3}\to 0 ,
\eeq
keeping $a_0 n^{1/3}$ fixed. Here, $\mu$ is the chemical potential and
$n$ is the density. Numerical results for the energy per particle
of dilute neutron matter are shown in Fig.~\ref{fig_latt}. We observe
that the results agree quite well with traditional many-body 
calculations. We also note that even with higher-order corrections
taken into account, the equation of state exhibits approximate 
universal behavior, with an effective $\xi\simeq (0.5\mbox{--}0.6)$.
For applications of the lattice method to
finite nuclei, see \cite{Borasoy:2006qn}.

\subsection{Perturbative EFT for Nuclear Matter}

The nuclear calculations discussed so far have all been nonperturbative.
However, renormalization group (RG) methods have been used to show
that the perturbativeness of internucleon interactions depends strongly on the
momentum cutoff and the density~\cite{Bogner:2003wn,Bogner:2006tw}.
Lowering the resolution via an RG evolution 
leaves observables and EFT truncation errors unchanged by construction
(up to approximation errors and omitted many-body contributions) but
shifts contributions between the potential and the
sums over intermediate states in loop integrals.
These shifts can weaken or even eliminate sources of non-perturbative
behavior such as strong short-range repulsion (\textit{e.g.}, from singular
chiral two-pion exchange) or the tensor force.
At sufficient density, effective range corrections and beyond 
suppress the effects of large $s$-wave scattering lengths~\cite{Bogner:2005sn}.

\begin{figure}[b!]
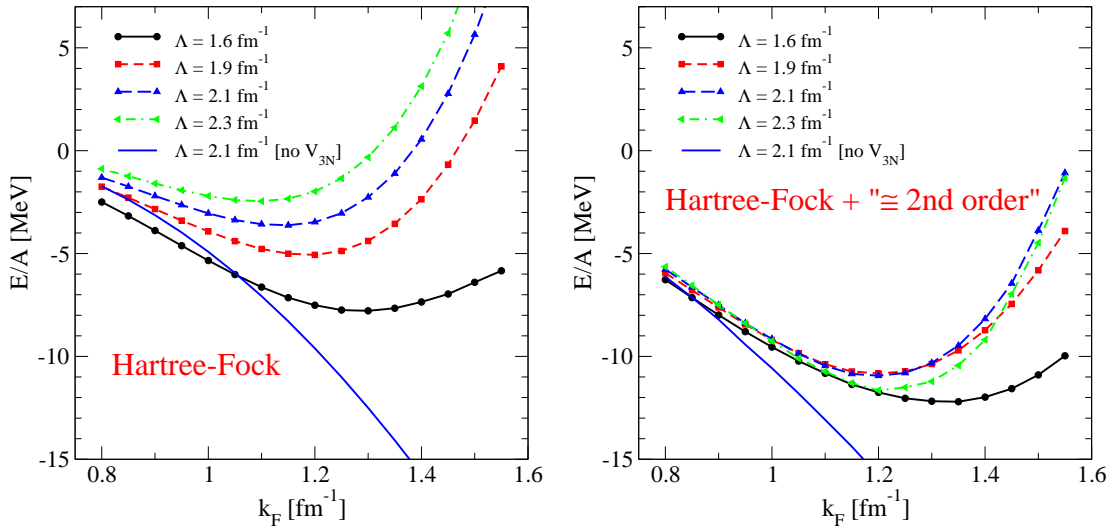

\centering
      \includegraphics*[width=2.8in]{paper_hf_2+3_rev4}%
      ~~~%
      \includegraphics*[width=2.8in]{paper_2ndorder_mstar_2+3_fullP_rev3}
 \caption{Nuclear matter energy per particle using
renormalization-group evolved low-momentum potentials with a range
 of cutoffs with 3NF fit to few-body
 binding energies~\cite{Bogner:2005sn}.  
 Left: Hartree-Fock.  Right: Hartree-Fock plus
 second order in the potential.}
 \label{fig:16}       
\end{figure}

As a consequence,
while nuclear matter is generally considered to be nonperturbative,
this is also resolution dependent.
Figure~\ref{fig:16} shows the energy
per particle in nuclear matter for several values of the RG cutoff
$\Lambda$ calculated in leading
order (Hartree-Fock) and second-order many-body perturbation 
theory~\cite{Bogner:2005sn}.
(Note: The initial potential used in these figures is not a chiral
EFT NN potential.  However, all NN potentials
fit to scattering phase shifts flow to very similar low-momentum potentials
by this range of cutoffs so similar results are expected.)
The three-body potential is of the N$^2$LO form, fit at each cutoff to the
binding energies of the triton and $^4$He.
Since the RG only changes short-distance physics, this is argued to be a good
approximation to the consistently evolved 3NF~\cite{Nogga:2004ab}. 

There are several encouraging features.
First, Hartree-Fock is a reasonable starting point for the description
of nuclear matter; this is patently false for conventional
phenomenological potentials, which do not even bind.
Second, the dependence on the cutoff is greatly reduced going to
second order.  Further calculations show that the third-order ladder
diagrams make a very small contribution~\cite{Bogner:2005sn}.
Third, with a fit to few-body properties, the minimum is reasonably close
to the empirical saturation point of (roughly) $-16\,$MeV per particle
with $\kf \approx 1.35\,\mbox{fm}^{-1}$
(indeed, the discrepancy is the order of uncertainties in the three-body
force).
These results motivate a program
to study nuclear matter with chiral EFT internucleon forces evolved to lower
resolution (which should also include studying unevolved chiral EFT
potentials fit with a lower cutoff).

The  increased perturbativeness in nuclear matter
with increased density and lower cutoff
can be understood physically from reduced phase space due to Pauli blocking and
the cutoff, combined with the favorable momentum dependence of the 
low-momentum interaction \cite{Bogner:2006tw,Bogner:2005sn}.
Pauli blocking means that particles with momenta below $\kf$ must forward
scatter (Hartree-Fock) or be excited out of the Fermi sea.
The latter amplitude is limited by the weakened coupling of occupied
and unoccupied states
that limits the volume of available momentum states (this is the phase
space restriction).
A consequence is that the  
saturation mechanism is now dominated by the three-body force contribution
(\textit{cf.} the NN-only curves in the figure),
rather than from the density dependence of two-body tensor
contributions.
For cutoffs in the range shown, the three-body contribution
still remains natural-sized according to chiral EFT power 
counting~\cite{Bogner:2005sn}, but is clearly quantitatively important.
The implication is that the 4NF contribution will also be important
at the level of about an MeV per particle at saturation, but this has yet
to be tested.

These results suggest that
an alternative EFT power counting may be 
appropriate at nuclear matter densities.
Kaiser and collaborators
have proposed a perturbative chiral EFT approach to nuclear matter
and then to finite nuclei through an energy
functional~\cite{Kaiser:2001jx,Kaiser:2002jz,Fritsch:2004nx}
(see also \cite{Lutz:1999vc}).
They consider Lagrangians both for nucleons and pions and
for nucleons, pions, and $\Delta$'s, and fit parameters to nuclear
saturation properties.  
They construct a loop expansion for the nuclear matter energy
per particle, which leads to an energy expansion of the form
\beq 
  E(\kf) = \sum_{n=2}^\infty \kf^n \, f_n(\kf/m_\pi,\Delta/m_\pi) \ ,
  \qquad [\Delta = M_\Delta - M_N \approx 300\,\mbox{MeV}]
\eeq
where each $f_n$ is determined from a finite number of
in-medium Feynman diagrams.
All powers of $\kf/m_\pi$ and $\Delta/m_\pi$ are kept in the $f_n$'s
because these ratios are not small quantities \cite{Kaiser:2006tu}.
A semi-quantitative description of nuclear matter is
found even with just the lowest two terms without 
$\Delta$'s and adding $\Delta$'s brings uniform improvement
(e.g., in the neutron matter equation of state).
There are open questions about power counting and convergence,
but many promising avenues to pursue.

\subsection{Density Functional Theory as an EFT}

Density functional theory (DFT)~\cite{DREIZLER90,ARGAMAN00,Fiolhais:2003}
is widely used in condensed matter and quantum chemistry to treat
large many-body systems.
It is based on the response of the ground-state 
energy to external perturbations of the density,
with fermion densities as the fundamental variables.
This means that the computational cost for DFT is far less than
for wave function methods and the calculations can be applied to heavy
nuclei. 
DFT is naturally formulated in an effective action framework~\cite{Polonyi2001}
and carried out
using an inversion method implemented with EFT power 
counting~\cite{FUKUDA94,VALIEV97,Furnstahl:2007xm}.

The simple prototype EFT from Section~\ref{sec:uniform} for a dilute system 
can be revisited in DFT by putting the fermions in a trap 
potential $v_{\rm ext}(\bfx)$ (\textit{e.g.},
a harmonic oscillator) and adding sources coupled to external 
densities~\cite{Puglia:2002vk}. 
Consider a single external source $J(\bfx)$ coupled to the density operator 
$\widehat \rho(x) \equiv \psi^\dagger(x)\psi(x)$ in the partition
function (neglecting normalization and factors of the temperature and volume
and suppressing $v_{\rm ext}$), 
\beq
    \mathcal{Z}[J] = 
    e^{-W[J]} \sim {\rm Tr\,} 
      e^{-\beta (\widehat H + J\,\widehat \rho) }
    \sim \int\!\mathcal{D}[\psi^\dagger]\mathcal{D}[\psi]
    \,e^{-\int\! [\mathcal{L} + J\,\psi^\dagger\psi]} 
    \;,
\eeq
with the Lagrangian from Section~\ref{subsec:prototype}.
The static density $\rho(\bfx)$ in the presence of $J(\bfx)$ is
\beq
  \rho(\bfx) \equiv \langle \widehat \rho(\bfx) \rangle_{J}
   = \frac{\delta W[J]}{\delta J(\bfx)}
   \;,
\eeq  
which we invert to find $J[\rho]$ and then Legendre transform from $J$ to
$\rho$:
\beq
   \Gamma[\rho] = - W[J] + \int\!d^3x\, J(\bfx) \rho(\bfx) 
%
\quad\mbox{with}\quad
   J(\bfx) = \frac{\delta \Gamma[\rho]}{\delta \rho(\bfx)}
   \longrightarrow 
   \left.
   \frac{\delta \Gamma[\rho]}{\delta \rho(\bfx)}\right|_{\rho_{\rm gs}(\bfx)
   } =0
   \;.
   \label{eq:Jofx}
\eeq 
For static $\rho(\bfx)$, $\Gamma[\rho]$ is proportional to 
the Hohenberg-Kohn energy functional, which
by Eq.~(\ref{eq:Jofx}) is extremized at the ground state density
$\rho_{\rm gs}(\bfx)$.

With $W[J]$ constructed as a diagrammatic expansion,  
EFT power counting gives us a means to invert 
from $W[J]$ to $\Gamma[\rho]$~\cite{FUKUDA94,VALIEV97}.  It proceeds by
substituting the decomposition 
$J(\bfx) = J_0(\bfx) + J_{1}(\bfx) + J_{2}(\bfx) + \ldots $
(where ``1'' means LO, ``2'' means NLO, and so on)
and corresponding expansions for $W$ and $\Gamma$ into Eq.~(\ref{eq:Jofx}) and
matching order-by-order with $\rho$ treated as order unity.
Here $J_0$ is chosen so that there are no corrections to
the zeroth order density at each order in the expansion;
the interpretation is that $J_0$ is the external potential that yields 
for a noninteracting system the exact density.
Zeroth order is the noninteracting system with potential $J_0(\bfx)$, 
 \beq
   \Gamma_0[\rho] = -W_0[J_0] + \int\!d^3x\, J_0(\bfx)\rho(\bfx) 
   \quad \Longrightarrow \quad \rho(\bfx)
      = \frac{\delta W_0[J_0]}{\delta J_0(\bfx)}  \; ,  
 \eeq     
which is the so-called Kohn-Sham system with the exact density!
To evaluate $W_0[J_0]$, we introduce orbitals $\{\psi_\alpha\}$ satisfying
(with $v_{\rm ext}$ made explicit)
\beq
       \bigl[ -\frac{{\nabla}^2}{2M}  +  \Vext({\bf x}) - J_0({\bf x})
       \bigr]\, \psi_\alpha(\bfx) = \varepsilon_\alpha \psi_\alpha(\bfx)
       \; ,
       \label{eq:psis}
\eeq
which diagonalize $W_0$, so that it yields a sum of $\varepsilon_\alpha$'s
for the occupied states.
We calculate the $W_i$'s and $\Gamma_i$'s up to a given order as
functionals of $J_0$ and then
determine $J_0$ for the ground state via a  self-consistency loop:
\beq
  { J_0} \rightarrow W_1 \rightarrow \Gamma_1 \rightarrow J_1
   \rightarrow W_2 \rightarrow \Gamma_2 \rightarrow \cdots
   \Longrightarrow
    \left.J_0(\bfx)\right|_{\rho = \rho_{\rm gs}} = 
      \left.\frac{\delta\Gamma_{\rm interacting}[\rho]}
        {\delta \rho(\bfx)}\right|_{\rho = \rho_{\rm gs}}
	 \; .    
\eeq
Adding sources coupled to other currents improves the functional variationally
and allows pairing to be treated within the same 
framework~\cite{Bhattacharyya:2004qm,Furnstahl:2006pa}.

\begin{figure}[t!]
\centering
      \includegraphics*[width=2.3in]{error_plot_tau_breakup140}%
      ~~~%
      \includegraphics*[width=3.0in]{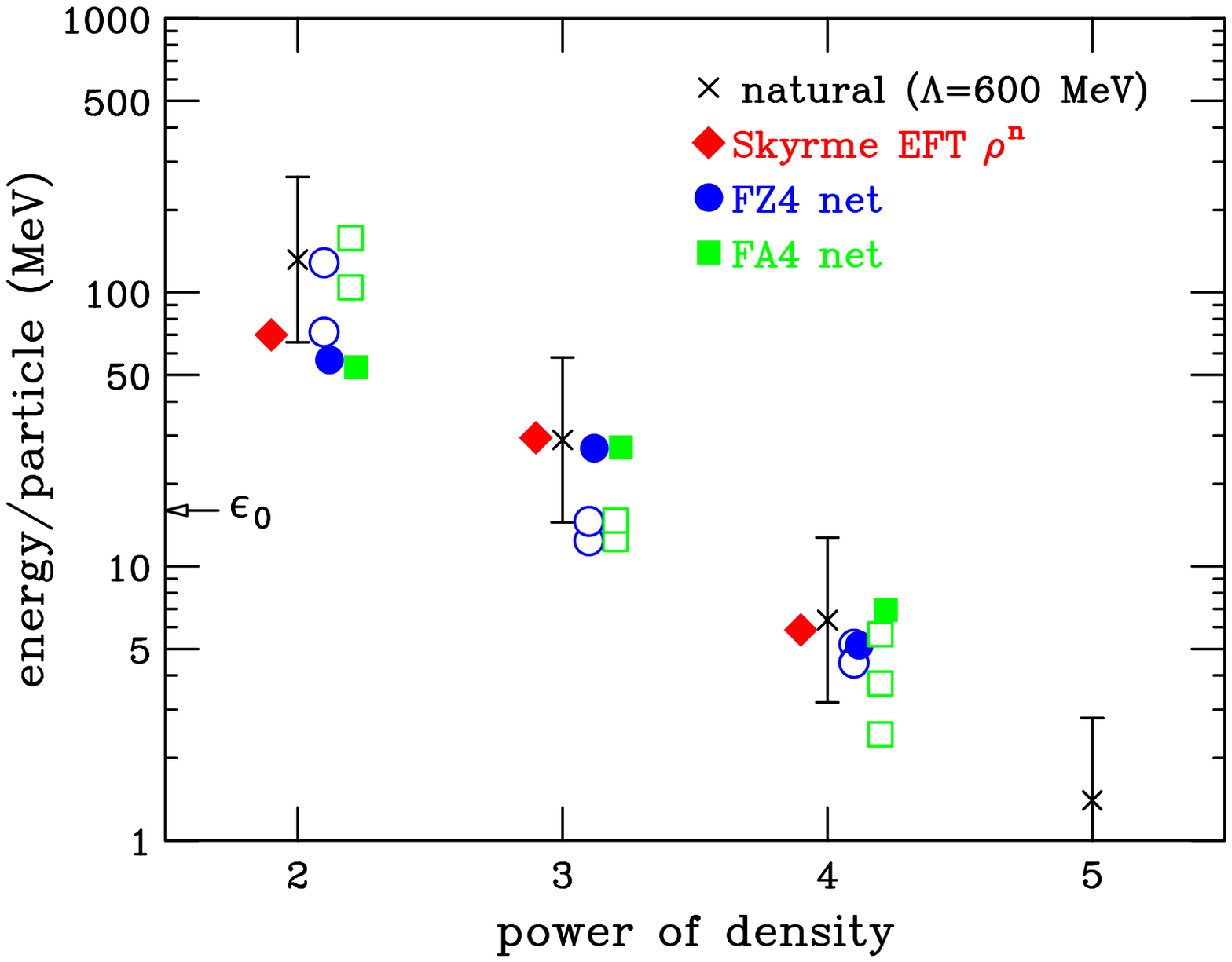}
 \caption{Estimates for energy functionals for a dilute
 fermions in a harmonic trap (left) and for three phenomenological
 energy functionals for nuclei (right).}
 \label{fig:estimates}       
\end{figure}

Figure~\ref{fig:estimates} shows how EFT power-counting estimates
predict the hierarchy of contributions to a DFT energy functional.
On the left are results for the energy per particle of 
$A=140$ fermions in a trap with short-range repulsive interactions.
The \textit{a priori}
estimates from terms at three different orders in the EFT
expansion 
(the counterparts to the terms in Eq.~(\ref{eq:Edensity}) plus
gradient corrections)
are shown with error bars that reflect a natural range for
the unknown coefficients (in this case from 1/2 to 2). 
These are compared to actual values, with good
agreement~\cite{Bhattacharyya:2004qm}. 
A similar exercise using a chiral-EFT-inspired power counting has been
applied to phenomenological nonrelativistic (Skyrme) and covariant
density functionals.  Results for terms organized by powers of the
density in each term 
are shown on the right in Fig.~\ref{fig:estimates}
and show that the predicted hierarchy
is realized~\cite{Furnstahl:2004xn,Furnstahl:2007xm}.  
 
The apparent success of many-body perturbation theory for nuclear
matter using low-momentum
potentials RG-evolved from chiral EFT input makes feasible
the construction of a nuclear DFT functional in the effective action
formalism that is compatible with nonrelativistic
Skyrme energy functional technology~\cite{Dobaczewski:2001ed,Bender:2003jk}.  
Work is underway as part of a large-scale five-year project to 
develop a universal nuclear
energy density functional (UNEDF) that will cover the entire table
of nuclides~\cite{unedf:2007}.
The goal is to generate systematically improved energy functionals
based on chiral EFT/RG input potentials, including theoretical error
estimates so that extrapolation to the driplines is under control.

The density matrix expansion (DME) of Negele and
Vautherin~\cite{Negele:1972zp,Negele:1975zz} 
has been extended
to three-body force contributions and applied in momentum space to provide
the first-generation functional~\cite{Furnstahl:2007xm,Bogner_dme}.
This construction is facilitated by analytic expressions for the long-range
pion contributions derived by Kaiser et al.~\cite{Kaiser:2002jz,Fritsch:2004nx}.
The functional
has the form of a generalized Skyrme functional with density-dependent
coefficients, including all allowed terms up to two derivatives, which 
means it can be directly incorporated into existing computer codes.
Cutoff
dependence can be used as a diagnostic tool for assessing missing elements of
the interaction, the many-body approximations, and the performance of the 
energy functional.
Benchmarking against NCSM and CC calculations for light- and medium-mass 
nuclei is possible by calculating the energy with an additional
external field, \textit{i.e.}, putting the nuclei in theoretically
adjustable traps.


\section{SUMMARY AND OUTLOOK}
  \label{sec:summary}

Effective field theory is a well-established technique with 
successes in all branches of physics.
Applications of EFT to finite density systems have many precursors
stretching back decades but implementations are relatively recent.
Many-body systems with short-range interactions are an ideal
testing ground for many-body EFT because of the universal nature
of the systems and the connection to experiment through cold atom
physics.

Far less developed is the application of EFT methods to nuclear
many-body systems.
The immediate impact of EFT on nuclear many-body calculations is
through the systematic organization of  effective Hamiltonians
for low-energy QCD using chiral effective field theory.
Of particular importance is the role of many-body forces.
We emphasize that while these Hamiltonians have many successes describing
scattering and properties of light nuclei, they are largely untested
at densities relevant for most nuclei and nuclear matter.  
Fortunately, computational tools such as the NCSM, CC, and lattice methods,
renormalization group techniques, and density functional theory
will funnel advances in chiral EFT to new predictions, so that
true tests are forthcoming.
More direct applications of EFT methods to many-body calculations are
in their infancy but there are clear incentives to pursue them.

This has necessarily been a shallow survey but the breadth of 
activity should be clear. Key developments are expected in the next
few years.  These include improvements to the chiral EFT potentials
such as full N$^3$LO three-body interactions and the corresponding
N$^3$LO Hamiltonian with $\Delta$ degrees of freedom and the subsequent
testing of power counting in light to medium mass systems.
In addition,
the consistent evolution of many-body forces with RG methods will
open the door to the full range of nuclei and nuclear matter.

Beyond the calculational tools,
effective field theory provides a new \emph{perspective} for nuclear many-body
calculations.
Whereas before one sought a universal Hamiltonian for all problem and energy
length scales, EFT exploits the infinite number of low-energy
potentials: rather than finding the ``best'' potential we use
a convenient or efficient one or work directly from a Lagrangian.
For a long time it was hoped that two-body data would be sufficient
for nuclear systems;
many-body forces were treated as a last resort, to be considered as an add-on.  
In EFT it is inevitable that many-body
forces and data are needed and they are directly tied to the two-body
interaction.
Before we would avoid divergences and hide them in form factors;
with EFT we confront and exploit them 
(\textit{e.g.}, using cutoff dependence as a tool).  
Finally, instead of choosing diagrams to sum by ``art,'' power counting
determines what to sum and establishes theoretical truncation errors.  

Many relevant and interesting topics were not considered here because
of space limitations.
Two major (related) areas largely unaddressed are the response to
external probes and nuclear reactions.
Another area is EFT at high temperature 
for many-body systems with large scattering length, which  
has been formulated using the virial
expansion~\cite{Bedaque:2002xy, Rupak:2006pu}
(see Refs.~\cite{Horowitz:2005nd, Horowitz:2005zv} for recent
applications of the virial expansion to hot dilute nuclear matter).
The EFT formulation of the finite temperature nuclear
many-body system with long-range pion interaction is a frontier.
Other nuclear systems where EFT can play a particular role are
hypernuclei~\cite{Nogga:2006ir} and halo nuclei~\cite{Bertulani:2002sz}.
Work to apply EFT methods to covariant hadronic field theories 
strives to understand the successes of relativistic mean-field
phenomenology~\cite{Furnstahl:2003cd}.
Finally, there is the challenge of making the connection
to lattice QCD~\cite{Savage:2006nv} (as opposed to EFT on the lattice).


\section*{Acknowledgments}
This work was supported in part by the National Science 
Foundation under Grant Nos.~PHY--0354916 and PHY--0653312, 
the Department of Energy under Grant No.~DE-FG02-03ER4126, and 
the UNEDF SciDAC Collaboration under DOE Grant 
DE-FC02-07ER41457.





\bibliographystyle{h-physrev3}
\bibliography{eftfds_posted}



\end{document}